\documentclass[10pt,a4paper]{article}
\usepackage{makeidx}
\usepackage{amssymb}
\usepackage{amsmath}
\usepackage{graphicx}
\usepackage{graphics}
\usepackage{times}
\usepackage[T1]{fontenc}
\usepackage[thinspace,thinqspace,squaren]{SIunits}
\usepackage{graphicx}
\usepackage{amsthm}
\usepackage{color}
\usepackage{hyperref}
\usepackage{multirow}

\usepackage{ctable}
\allowdisplaybreaks

\usepackage{algorithm}
\usepackage{algorithmic}

\twocolumn

\begin{document}
\date{}

\title{A Small-Scale Prototype to Study the Take-Off of Tethered Rigid Aircrafts for Airborne Wind Energy}


\author{Lorenzo~Fagiano,
        Eric Nguyen-Van,
        Felix Rager,
        Stephan Schnez,
        and Christian Ohler%
\thanks{This is the pre-print of a paper submitted for publication. The authors are with ABB Switzerland Ltd., Corporate Research, 5405 Baden-D\"{a}ttwil - Switzerland. E-mail addresses: \{lorenzo.fagiano $|$ eric-nguyen-van $|$ felix.rager $|$ stephan.schnez $|$ christian.ohler\}@ch.abb.com.}
\thanks{Corresponding author: Lorenzo Fagiano.}}
\maketitle

\begin{abstract}
The design of a prototype to carry out take-off and flight tests with tethered aircrafts is presented. The system features a ground station  equipped with a winch and a linear motion system. The motion of these two components is regulated by an automatic control system, whose goal is to accelerate a tethered aircraft to take-off speed using the linear motion system, while reeling-out the tether from the winch with low pulling force and avoiding entanglement. The mechanical, electrical, measurement and control aspects of the prototype are described in detail. Experimental results with a manually-piloted aircraft are presented, showing a good matching with previous theoretical findings.
\end{abstract}


%

\section{Introduction}\label{S:introduction}

The term Airborne Wind Energy (AWE) refers to a series of technologies to convert wind energy into electricity by using aircrafts tethered to the ground, \cite{Ahrens2013,Fagiano2012}. The claimed advantages of AWE systems over conventional wind turbines are, for the same rated power, a lower capital cost, since no expensive structure like the tower has to be built, and a higher energy yield, thanks to the possibility to harvest wind energy at higher altitude (usually in the range 200-400 m above ground), where the time-share of above-rated wind speeds is larger. The main disadvantages are a higher complexity, due to the heavy reliance on active control systems to achieve stable and optimal operation, and higher operation and maintenance costs, caused by the use of materials, in particular for the tether, which are not as durable as those of conventional wind generators. AWE concepts have been first envisioned in the late '70s \cite{Manalis1976,Loyd1980}, however their technical development started only in the early 2000s and has seen a quite steady progress in the last decade. Today, besides few notable exceptions \cite{Vermillion2014}, most of the approaches under development exploit the so-called crosswind motion, i.e. a flight trajectory roughly perpendicular to the wind performed at relatively large speed, and either onboard power conversion \cite{VanderLind2013} or ground-based one \cite{Ruiterkamp2013,Bormann2013,Vlugt2013,Erhard2015}. 

In the scientific literature, systems with ground-based generation and soft kites are by far the ones that received the largest attention, with several contributions concerned with aerodynamics \cite{Breukels2010a,Bosch2014,Leloup2014,Terink2011,Breukels2014} and controls \cite{Canale2010,Ilzhoefer2007,Fagiano2014,Zgraggen2016,Erhard2012a,Erhard2015}. On the other hand, fewer results have been published pertaining to systems with  ground-based electricity generation and rigid aircrafts\cite{Ruiterkamp2013,Stuyts2015}, and even fewer for system with onboard generators \cite{VanderLind2013}. One cause of such a disparity is the fact that concepts with rigid aircrafts are inherently more complex than those based on soft kites, hence more difficult to study within academia, especially when it comes to experiments. In fact, with power kites several degrees of freedom are constrained by the bridle design, so that two control inputs, usually a steering deviation and the force exerted on the tether, are enough to obtain stable flight patterns and produce energy. When two or more tethers are present, control can even be achieved using ground-based actuators and sensors only \cite{Fagiano2014,Zgraggen2016}. With rigid aircrafts, on the contrary, an onboard autopilot has to be developed, which is able to coordinate with the ground station in order to actively stabilize all of the degrees of freedom of the system. Considering a standard aircraft design, like a propelled glider, this entails the coordination of five onboard control inputs (the aerodynamic control surfaces and the propeller) and one on the ground (the tether force). Moreover, the superior aerodynamic performance of rigid aircrafts leads to faster system's dynamics, which render the control problem more challenging, in particular for what concerns the interaction with the tether. Finally, soft kites are much more resilient to impacts and easier to repair than rigid aircrafts, hence making experimental tests with the latter even more challenging and time-consuming.

With the aim of easing the mentioned difficulties, the first contribution of this paper is the description of a small-scale, low-cost prototype, which can be used to study experimentally some relevant aspects of  AWE systems with ground-based generation and rigid aircrafts. Similar in spirit to \cite{Fagiano2015}, concerned with a small-scale prototype to control soft kites, we present the design of the hardware, software and operation of the prototype with enough details to allow other teams of researchers to replicate and/or improve the setup.

The operation of any AWE system can be divided in three main phases, which shall be carried out autonomously: take-off, power generation, and landing. Considering the current development status of AWE systems with ground-based generation, autonomous power generation has been assessed both theoretically and experimentally and several related contributions are available in the scientific literature, see e.g. \cite{Erhard2015,Ruiterkamp2013,Milanese2014,Vlugt2014}. The same cannot be said about autonomous take-off and landing of the aircraft. These two topics have been explored by few papers in the scientific literature, mainly with theoretical and numerical analyses only, and demonstrated by AWE companies only to a limited extent. Actually, the capability to carry out take-off and landing in compact space and in an economical way is one of the main technical challenges (hence risks) still standing in AWE development. Again, this holds particularly for systems using rigid aircrafts, for which there is evidence of autonomous take-off \cite{ampyx}, however by using a winch launch that requires
a significant space in all directions in
order to adapt to different wind directions. In the scientific literature, \cite{ZaGD13} presents a simulation study for a rotational take-off, while in \cite{Bont10}  an analysis of several approaches is carried out and three alternatives are deemed the most promising: buoyant systems, linear ground acceleration plus on-board propeller, and rotational take-off. Then, the rotational take-off is examined in more detail by means of numerical simulations. In \cite{Fagianoa}, a theoretical analysis is presented, which shows how a linear take-off approach appears to be the most viable one, according to different performance criteria. About the landing, to the best of the authors' knowledge there is only one contribution in the literature \cite{Nguyen-Van2016}, which studies this aspect through numerical simulations after presenting a possible control approach for the aircraft to realize a cycle of tethered take-off, low-tension flight and landing.
\begin{figure*}[!htb]
 \begin{center}
  \includegraphics[trim= 0cm 0cm 0cm 0cm,width=2\columnwidth]{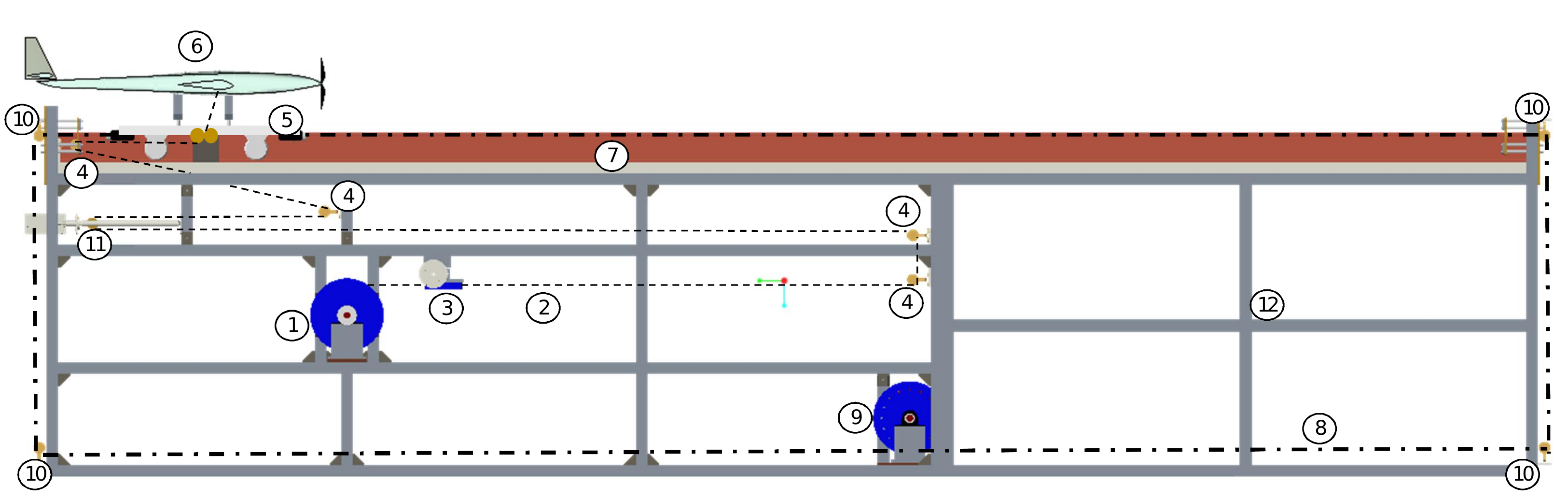}
  \caption{Rendering of the small-scale prototype built at ABB Corporate Research. The numbers in the picture indicate: 1. the winch, 2. the tether connected to the aircraft (dashed line), 3. the spooling mechanism, 4. the series of pulleys that redirect the tether from the winch to the aircraft, 5. the slide, 6. the aircraft, 7. the rails, 8. the tether used to pull the slide in backward/forward directions (``slide tether'', dash-dotted line), 9. the drum hosting the slide tether, 10. the pulleys that redirect the slide tether, 11. the mass-spring system, 12. the frame. As a reference, the rails' length in this rendering is 5.2$\,$m.}
  \label{F:sys_sketch}
 \end{center}
\end{figure*}

The second contribution of this paper is to demonstrate the take-off in a compact space and the subsequent flight of a tethered aircraft, using the described small-scale prototype. In particular, the aircraft is manually piloted, while the ground station is fully autonomous. We describe the automatic control system for the ground station, which regulates the linear acceleration of the aircraft up to take-off speed and the tether reeling during take-off and flight. We present experimental results collected by a series of sensors on the ground and onboard, and comment on the matching between such results and the theoretical ones derived in \cite{Fagianoa}. 

Together, the two mentioned contributions (description of the prototype and demonstration of tethered take-off and flight) make it possible for other research teams to realize a similar system and use it to  investigate a number of possible research topics, from aircraft design and control to filtering and state estimation. As an example, we used this prototype to develop an onboard control system for the aircraft, able to achieve fully autonomous take-off and flight. A movie of the related experimental results is available online \cite{Fagiano2016a}. The system modeling, identification and controller design aspects related to such an autopilot, as well as the obtained experimental results, fall beyond the scope of the present paper, whose focus is on the prototype hardware and ground station control. The autonomous take-off and flight of the aircraft are therefore subject of a separate contribution \cite{Fagiano2016submitteda}.


The paper is organized as follows. Section \ref{S:proto_layout} describes the layout of the prototype and its operation. Section \ref{S:design} provides details on the design of the prototype, with particular care to the most critical components. Section \ref{S:control_design} is concerned with the automatic control design of the ground station. Section \ref{S:results} presents the experimental results and concluding remarks are given in Section \ref{S:conclusions}.

\section{Prototype layout and principle of operation}\label{S:proto_layout}

\begin{figure}[!hbt]
 \begin{center}
  \includegraphics[trim= 0cm 0cm 0cm 0cm,width=\columnwidth]{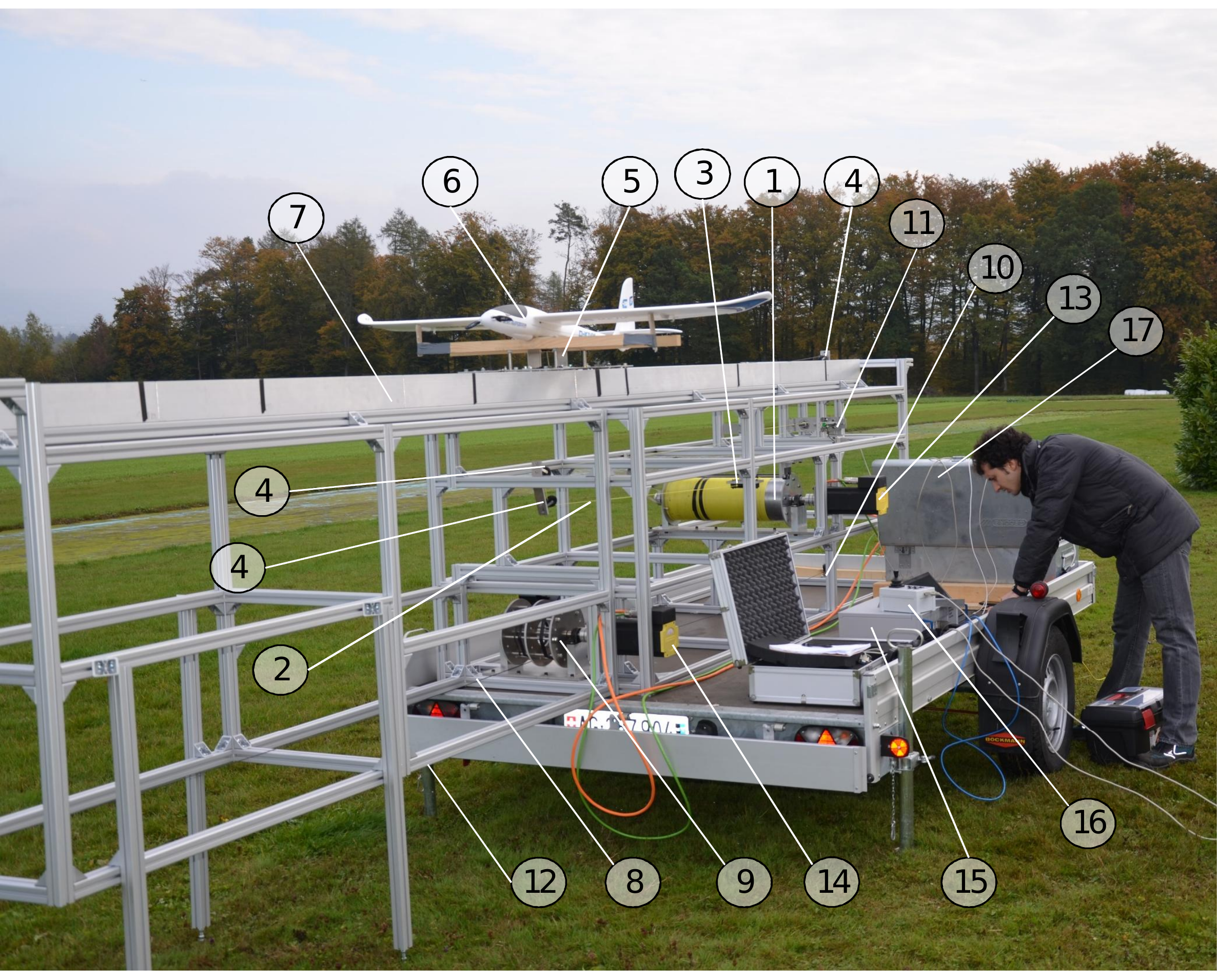}
  \caption{Picture of the small-scale prototype built at ABB Corporate Research. The numbers 1. to 12. correspond to the components described in the caption of Fig. \ref{F:sys_sketch}. In addition, the picture shows: 13. the winch motor, 14. the slide motor, 15. the real-time machine, 16. the human-machine interface, 17. the metal enclosure containing the power supply and motor drives.}
  \label{F:sys_pic}
 \end{center}
\end{figure}

\begin{figure*}[!hbt]
 \begin{center}
  \includegraphics[trim= 0cm 0cm 0cm 0cm,width=1.6\columnwidth]{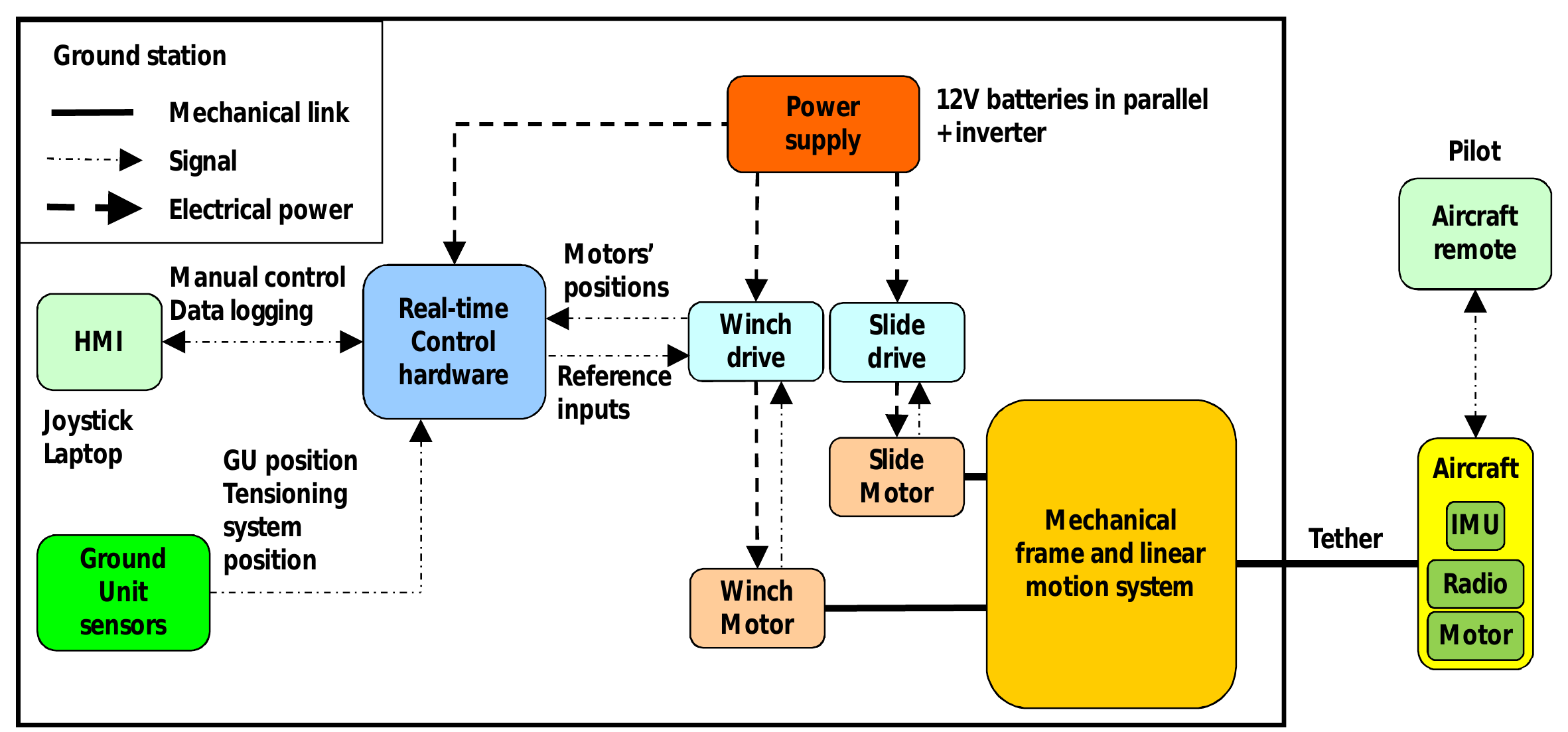}
  \caption{Conceptual layout of the prototype showing the various sub-systems and their mechanical and electrical (power and signal) interconnections.}
  \label{F:sys_concept}
 \end{center}
\end{figure*}

A rendering of our small-scale prototype is shown in Fig. \ref{F:sys_sketch}, highlighting all the main mechanical components. Moreover, Fig. \ref{F:sys_pic} presents a picture of the realized prototype, and Fig. \ref{F:sys_concept} a conceptual layout of the various subsystems and their mechanical and electrical links, the latter divided into power and signal. The system under consideration is composed of two main sub-systems, the ground station and the aircraft, connected by a tether. The ground station includes a mechanical frame supporting a number of components needed to achieve two main tasks: accelerating the aircraft from standstill to take-off speed, and controlling the tether reeling in order to limit the pulling force while, at the same time, avoiding entanglement and excessive sag.
In particular, referring to Figs. \ref{F:sys_sketch}-\ref{F:sys_pic}, the tether is coiled around a winch and passes through a series of pulleys before attaching to the aircraft. The winch is connected to a spooling system that translates backward/forward as the tether is reeled in/out, in order to distribute the latter evenly along the winch's axial length. One of the pulleys which the tether passes through is installed on a moving plate, connected to a spring. This component is used to reduce the stiffness of the link between the ground station and the aircraft and to control the winch speed, as described in section \ref{S:control_design}. Before the take-off, the aircraft is installed on a slide, able to move on rails and controlled by a linear motion system, which in our prototype is composed of a second tether wound around a drum (``slide drum'') and connected to both ends of the slide via another series of pulleys. The winch and the slide drum are each connected to an electric motor, controlled by a drive. A supply module provides the electrical power required by the drives and by the control hardware. The latter acquires the motors' position measurements from the drives and the spring's compression from a linear potentiometer and computes the reference position and speed values for the slide and winch motors, respectively (see Fig. \ref{F:sys_concept}). A human-machine interface (HMI) allows the operator to interact with the ground station. Regarding the aircraft, we consider a conventional airplane design whose main functional components, for the sake of our application, are a tether attachment and release mechanism, a front propeller, the typical control surfaces (ailerons, elevator, rudder and flaps), an onboard electronic control unit, a radio receiver linked to a remote controller, employed by a pilot on ground, finally onboard position, inertial, attitude and airspeed sensors.

The operation of the described system is quite straightforward: when the control system of the ground station receives a ``take-off'' command (in our prototype issued by a human operator, in an eventual final application originated by a supervisory controller e.g. on the basis of the wind conditions), it accelerates the aircraft up to take-off speed using about half of the rails' length, the second half being needed to brake the slide. At the same time, the winch accelerates in a synchronized way with the slide, in order to reel-out the tether at just the right speed to avoid pulling on the aircraft (which would quickly cause a stall condition) while also avoiding to entangle the tether due to an excessive reel-out with no pulling force. This is achieved by a feedforward/feedback control strategy described in detail in section \ref{S:control_design}. The aircraft has also to coordinate with the ground station, in such a way that the onboard motor is ramped up to full power in order to quickly climb to a safe altitude. As anticipated, in this paper we assume that a human pilot controls the aircraft and carries out such a  maneuver. In \cite{Fagiano2016submitteda} we describe an automatic controller of the aircraft, which synchronizes with the ground station using the longitudinal acceleration measured onboard. After the take-off, the pilot is in charge of maneuvering the aircraft to fly in a bounded region relatively close to the ground station, in order not to exceed the total length of tether installed on the winch. Typically flown patterns are figures-of-eight or ellipsoidal trajectories roughly above the ground station. In this phase, the winch controller has to still regulate the reeling speed to fulfil the same two conflicting objectives (low pulling force and low tether sag) while the aircraft periodically flies closer and farther from the ground station.

In the next section, we provide more details about each of the mentioned sub-systems and components, as well as design guidelines and the specific design choices we made for our small-scale prototype.

\section{Prototype design}\label{S:design}

\subsection{Ground station}\label{SS:ground_station_design}
\subsubsection{Mechanics}\label{SSS:mechanics_design}
The mechanical frame has to support all other components and withstand the forces exerted by the aircraft through the tether, as well as the torques (motor, inertia and friction) acting on the winch and on the slide drums. Equations to compute the tether's force as a function of the aircraft's parameters during crosswind flight can be found in several references \cite{Loyd80,Ahrens2013,Fagiano2012,Fagiano2015}. Dimensioning the frame according to these equations provides large enough resistance for all types of loads that can be expected during take-off and low-tension flight, since during these maneuvers the tether force is much lower than in crosswind conditions. In our design, we actually dimensioned all of the components such that the weakest element in the mechanical system is the aircraft, which is also the cheapest component since we employed commercially available model gliders made of styrofoam. Other important aspects in the frame design for a research prototype are transportability, modularity and the possibility to achieve a roughly horizontal orientation of the rails also on uneven terrain. In our prototype, we employed T-slotted aluminum framing profiles forming a structure that can be easily modified. The frame is divided in two halves, which we loaded side by side on a trailer for transportation, and bolted head-to-head for testing. We used four extensible profiles to support the part of the frame overhanging the trailer and to easily adapt to rough terrains, see Fig. \ref{F:sys_pic}. The final dimensions of the frame we built are approximately 1m$\times$1m$\times$3m (width-height-length) for the part bolted to the trailer, and 0.6m$\times$1m$\times$2.4m for the overhung part, thus providing a total rail length of about 5.4$\,$m. The frame is made of 0.04$\times$0.04 aluminum profiles and it features three
``layers'', see Fig. \ref{F:sys_sketch}: the lower one for the slide drum and tether, the middle one for the winch, and the upper one for the mass-spring system. Finally, the rails are mounted on top. We adopted such a layered arrangement to provide enough room for modifications during the experimental activities, in order to cope with the uncertainty associated with the research activity. A much more compact arrangement can be easily achieved in an eventual final design, for example with all the components on just one layer and packed much closer one to the other.

\begin{figure}[!hbt]
 \begin{center}
  \includegraphics[trim= 0cm 0cm 0cm 0cm,width=0.85\columnwidth]{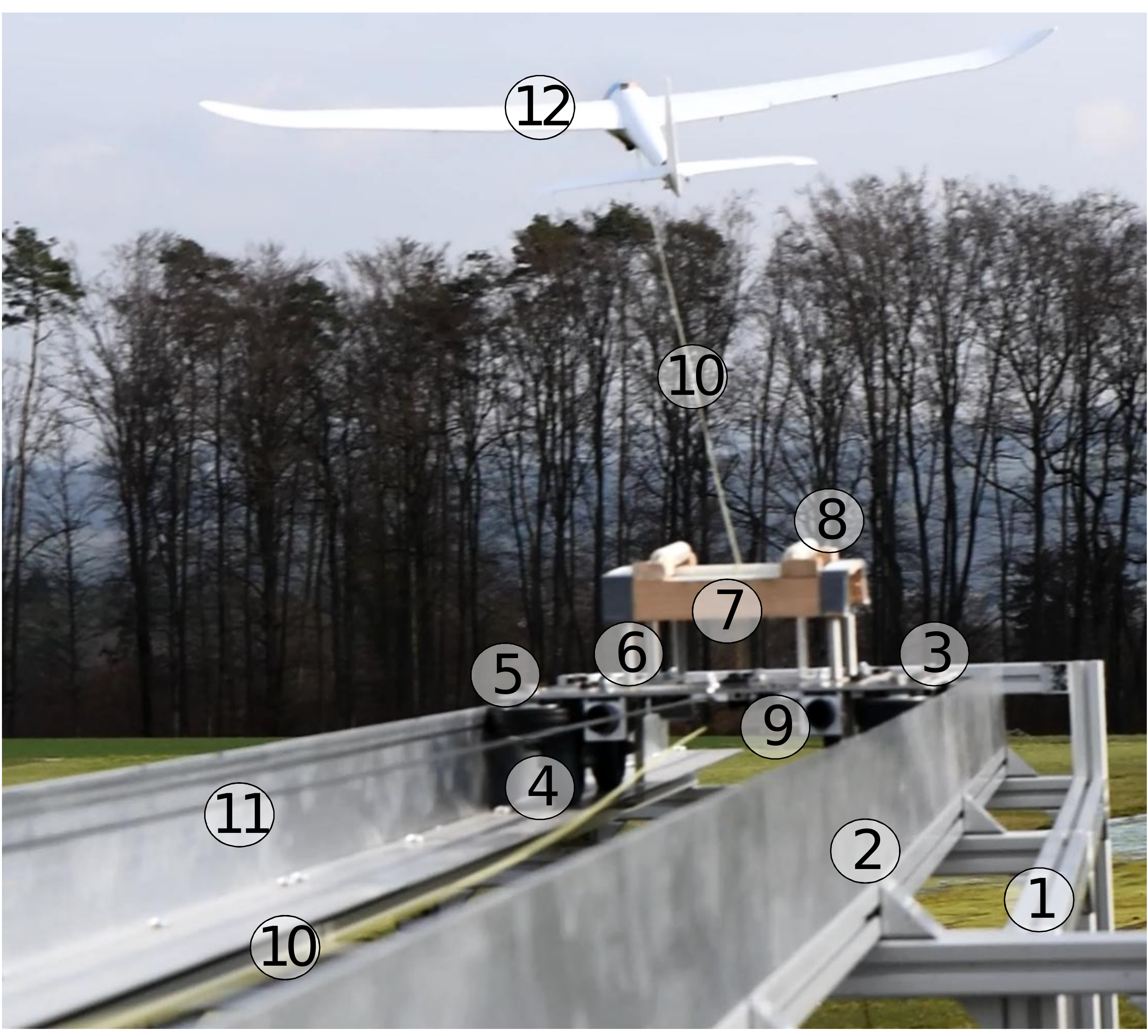}
  \caption{Snapshot of the rails, slide and aircraft shortly after take-off. The picture shows: 1. the frame, 2. the rails, 3. the slide floor, 4. the slide wheels, 5. the vertical-axis wheels, 6. adjustable steel shafts, 7. wooden supports for the aircraft, 8. plastic rollers, 9. bumpers, 10. tether connected to the aircraft, 11. slide tether, 12. aircraft.}
  \label{F:rails_slide}
 \end{center}
\end{figure}

For the rails, we employed L-shaped aluminum profiles bolted to the frame, on which the slide can roll thanks to four rubber wheels, see Fig. \ref{F:rails_slide}. The slide was designed in order to easily change the initial aircraft pitch. In particular, we employed two wooden bars mounted on the slide floor via adjustable shafts are used to support the plane and can be tuned in height and distance, in order to set the starting pitch angle. The wooden supports feature plastic rollers with bearings, which allow the tether to slide with very small friction. We arranged four such rollers in order to cope with all the possible directions that the tether can take during flight, i.e. 360$^\circ$. The tether passes through the slide floor by means of an in-line exit block with two low-friction nylon pulleys, of the kind used for sailing boats, see e.g. \cite{Harken}. Finally, in order to avoid lateral oscillations during the take-off, we installed four vertical-axis rubber wheels at the slide's corners. To limit its mass, the slide floor has to be as thin as possible compatibly with the need to support all the mentioned components. In our design we used a 0.4m$\times$0.6m$\times$0.004m aluminum plate. The slide's length has to be subtracted from the rail's length when computing the distance available for the take-off, which in our case reduces to about 4.8$\,$m. The slide floor, wooden supports, rollers and vertical-axis wheels are visible in Fig. \ref{F:rails_slide} as well. The final mass of the slide in our design is about 10$\,$kg.

\begin{figure}[!hbt]
 \begin{center}
  \includegraphics[trim= 0cm 0cm 0cm 0cm,width=0.85\columnwidth]{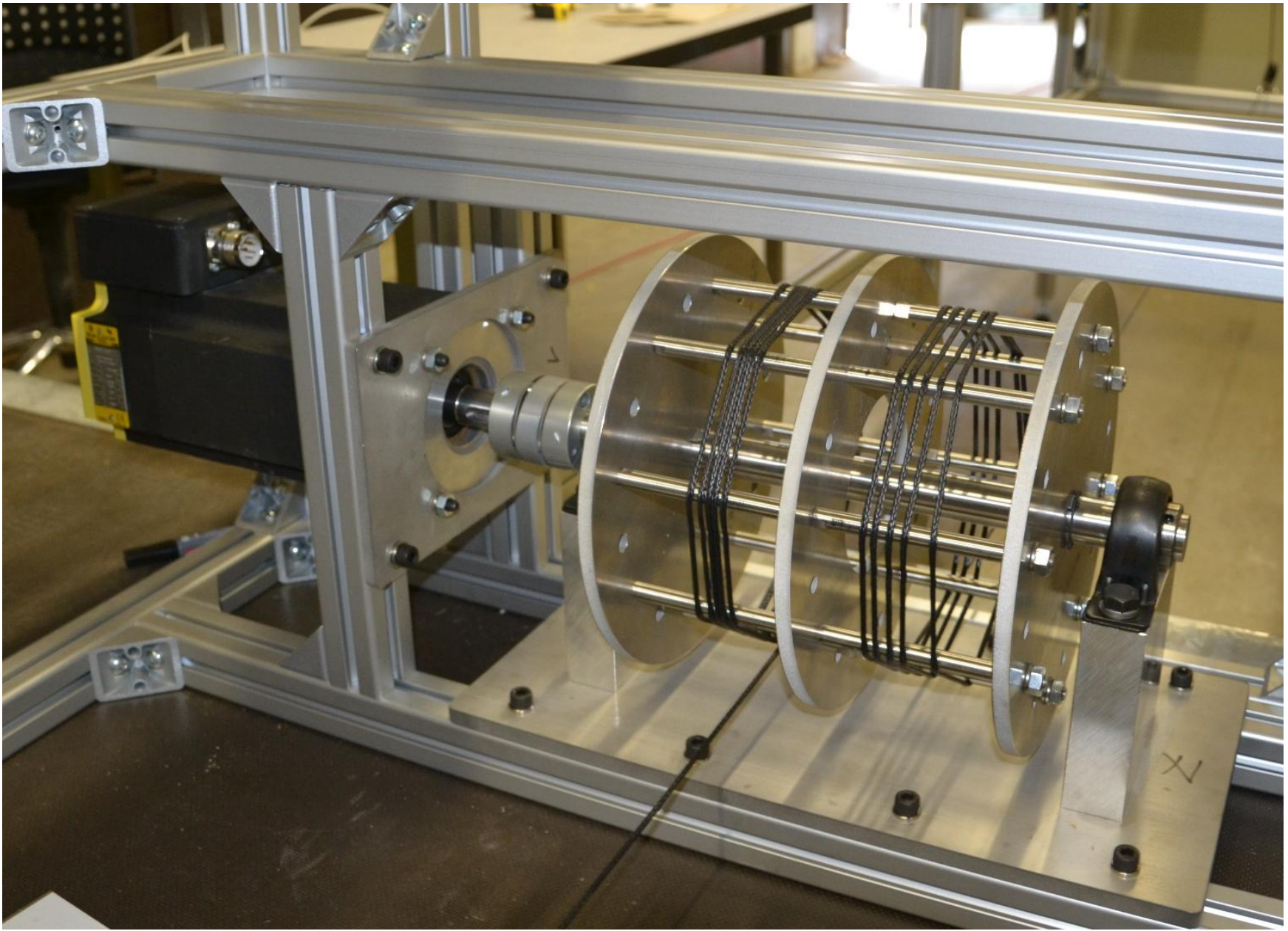}
  \caption{Picture of the slide drum with the tether installed and of the slide motor.}
  \label{F:slide_drum}
 \end{center}
\end{figure}

As mentioned, in our prototype the slide is pulled in both directions by means of a tether attached to the two sides of the slide's floor, see again Fig. \ref{F:rails_slide}. We employed a 0.003m-diameter ultra-high-molecular-weight polyethylene (UHMWPE) tether for this purpose, with a minimum breaking load of 10$^4\,$N which is much larger than the maximum force acting on the slide tether, of the order of few hundreds of N. The slide drum shall be as light as possible and be able to reel-in the slide tether on one side while at the same time reeling-out on the other side. To achieve these features, in our design we used three aluminum disks connected by steel rods bolted with regular spacing around a circular pattern with 0.1m of radius, see Fig. \ref{F:slide_drum}. The aluminum disks create two half-drums, each one 0.1m-wide. The drum is linked to the slide motor via a joint that we sized according to the maximum motor torque of $26\,$Nm.

\begin{figure}[!hbt]
 \begin{center}
  \includegraphics[trim= 0cm 0cm 0cm 0cm,width=0.85\columnwidth]{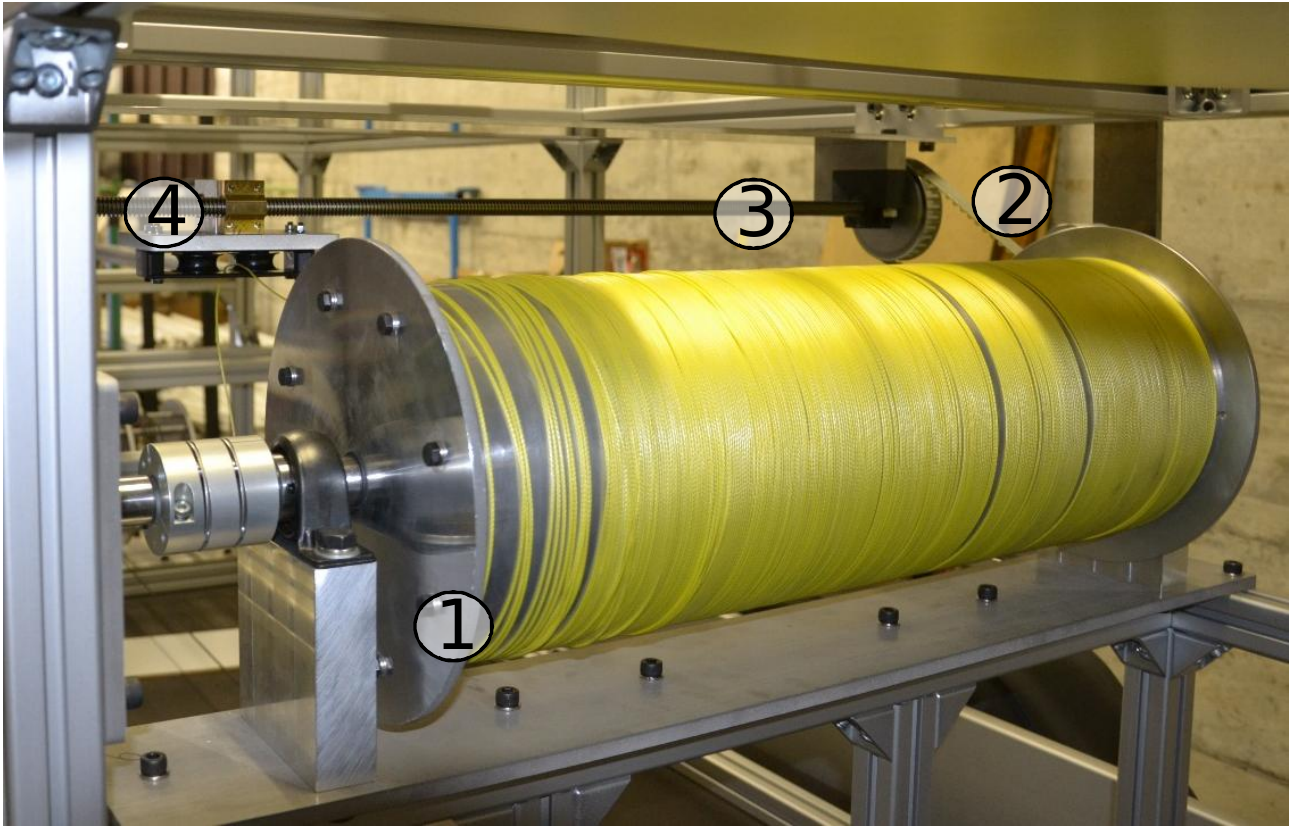}
  \caption{Picture of 1. the winch, 2. the belt transmission system, 3. the leadscrew and 4. the moving organizer block.}
  \label{F:winch_leadscrew}
 \end{center}
\end{figure}

The aircraft's tether shall be chosen with the smallest diameter possible compatibly with the expected loads, in order to limit the tether drag. The  material employed by most research groups and companies in AWE is UHMWPE, for its large minimum breaking load and lightweight. We employed a 0.002m-diameter UHMWPE tether for the prototype, with a total length of 150m. The minimum breaking load is about 4,500N, way above the breaking load of the attachment mechanism on the aircraft. Regarding the winch, its main task is to withstand the tether force and to reel-out/in the tether fast enough to coordinate with the aircraft movement, while avoiding tether entanglement. In our prototype we manufactured a rather bulky drum of aluminum with a thickness of 0.01$\,$m, a radius of 0.1$\,$m and an axial length of 0.6$\,$m. Considering the 0.002$\,$m of tether diameter, this corresponds to about 190$\,$m of tether that could be stored in a single layer. To control the tether spooling without the need of additional servomotors, we designed a linear motion system composed by a leadscrew driven by the winch through a belt transmission system, see Fig. \ref{F:winch_leadscrew}. The rotation of the leadscrew translates a small carriage where we mounted an organizer block, composed of two nylon pulleys \cite{Harken} which the tether passes through. We sized the leadscrew and the transmission ratio of the belt such that a full revolution of the winch corresponds to 0.002$\,$m (i.e. equal to the tether diameter) of translation of the spooling block, hence theoretically leaving no gap between two consecutive turns of the tether on the drum.

\begin{figure}[!hbt]
 \begin{center}
  \includegraphics[trim= 0cm 0cm 0cm 0cm,width=0.85\columnwidth]{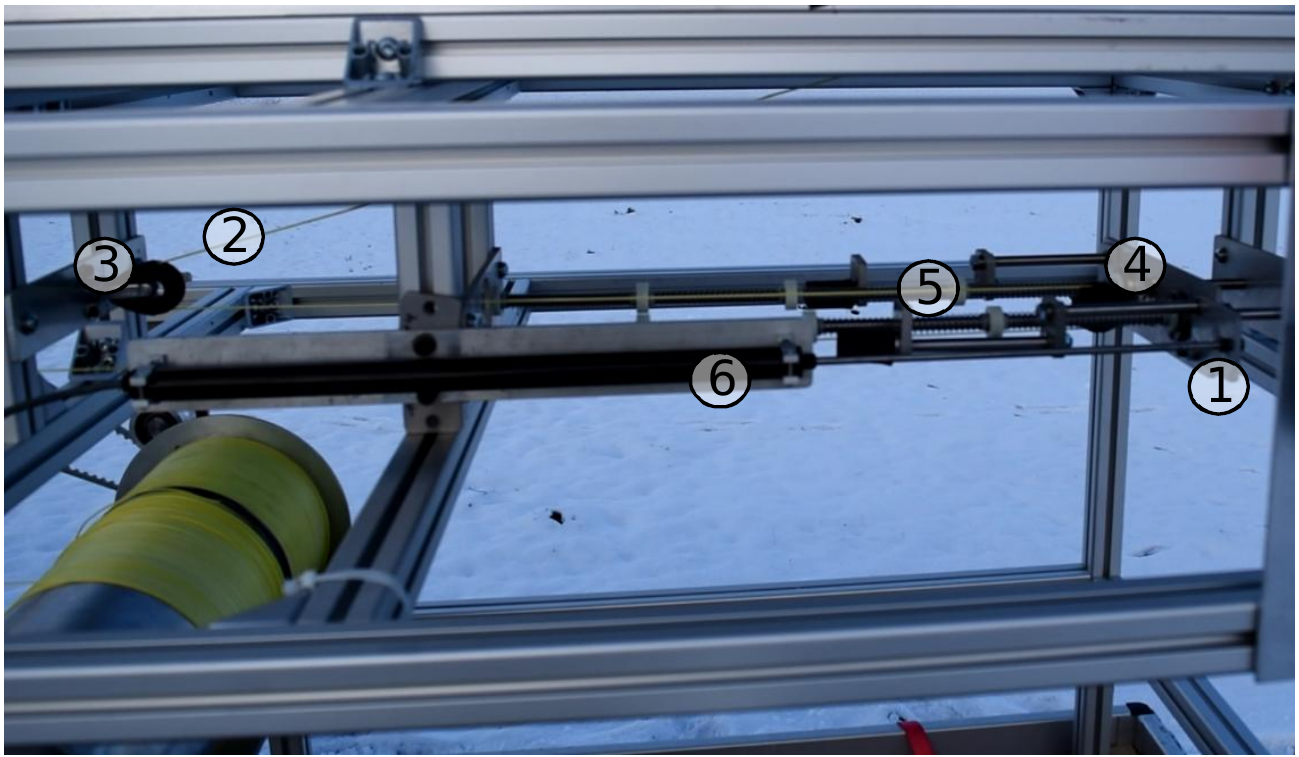}
  \caption{Picture of the mass-spring system. The tether 2. passes through a stand-up pulley 4., attached to a mobile plate 1. which can move on steel shafts, each one supporting four springs in series, 5.. A linear potentiometer 6. is linked to the mobile plate and measures its displacement, i.e. the spring compression. One of the fixed stand-up pulleys 3. used to re-direct the aircraft tether is also shown.}
  \label{F:tensioning_system}
 \end{center}
\end{figure}

Regarding the mass-spring system, Fig. \ref{F:tensioning_system} provides a picture and description of the solution we built in our prototype. The main task of this system is to limit the pulling force exerted by the tether on the aircraft, providing enough time for the winch control system to react and increase the reel-out speed until the tether is not taut anymore. To achieve this goal, the spring stiffness and travel need to be properly designed, moreover the spring position is measured and employed as feedback variable by the winch controller. The latter is described in detail in section \ref{S:control_design}, while in the following we introduce an approach to size the spring travel and stiffness as a function of the aircraft and winch features, by means of a simple numerical simulation. Indeed, the mass-spring system proved to be the most critical component in the whole mechanical design of the prototype. If the travel and stiffness of the spring are not well-designed, tethered flight might become impossible because when the tether is taut and the spring is fully compressed the pulling force increases very quickly, due to the high stiffness of UHMWPE, and there is not enough time for the winch to react before stalling the aircraft. This holds in particular when the tether is roughly aligned with the aircraft's longitudinal body axis and pulls backwards, for example right after take-off (while on the contrary in crosswind conditions the tether does not represent a problem, rather it's required to achieve high speed and generate power according to the well-known equations developed in \cite{Loyd80}).

\begin{figure}[!hbt]
 \begin{center}
  \includegraphics[trim= 0cm 0cm 0cm 0cm,width=0.85\columnwidth]{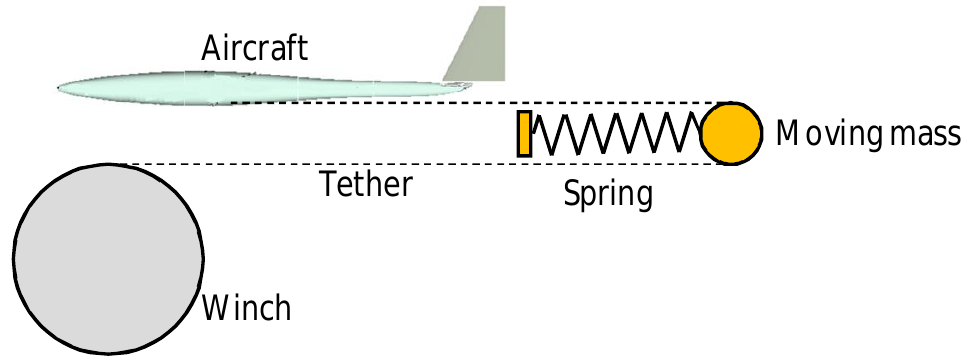}
  \caption{Simple model to study the interactions between aircraft, mass-spring system and winch in order to design the spring's travel and stiffness.}
  \label{F:spring_calc_setup}
 \end{center}
\end{figure}

We consider a simplified setup shown in Fig. \ref{F:spring_calc_setup}, where the aircraft moves along one dimension and the tether force is aligned with the drag force.
We consider that no wind is present, that the angle of attack of the aircraft is constant (hence the drag coefficient $C_D$ is fixed) and that the onboard propeller is providing the maximum thrust, $\overline{T}$. Finally we assume that the tether behaves like a spring whose stiffness $K_\text{t}(t)$ depends on the tether length $l(t)$, where $t$ is the continuous time variable:
\begin{equation}\label{E:tether_stiffness}
K_\text{t}(t)=\dfrac{\overline{F}_\text{t}}{\overline{\varepsilon}_\text{t}\,l(t)},
\end{equation}
where $\overline{F}_\text{t}$ is the breaking load and $\overline{\varepsilon}_\text{t}$ the corresponding elongation. The equation of motion of the aircraft in such a simplified setup is then:
\begin{equation}\label{E:aircraft_simple}
\ddot{x}(t)=\overline{T}-\dfrac{\rho\, C_D\, A\, \dot{x}^2(t)}{2m}-\dfrac{\text{sat}\left(K_\text{t}(t)(x(t)-l(t)),0\right)}{m},
\end{equation}
where $x$ is the aircraft position ($\dot{x}\doteq\frac{dx}{dt}$), $x(t)-l(t)$ is the tether elongation, $\rho$ is the air density, $A$ is the aircraft's effective area, $m$ its mass and $\text{sat}\left(F,0\right)$ is a saturation function that returns 0 if $F<0$, otherwise $F$. The tether length, for the sake of computing its elongation, is the sum of two contributions: the amount that was reeled-out from the winch, and the one yielded by the spring's compression:
\begin{equation}\label{E:tether_length}
l(t)=R_\text{w}\theta_\text{w}(t)+2x_\text{s}(t)
\end{equation}
where $R_\text{w}$ is the winch radius, $\theta_\text{w}$ its angular position, and $x_\text{s}(t)$ is the spring displacement. The dynamic behavior of the latter is governed by the following equation:
\begin{equation}\label{E:spring_simple}
\ddot{x}_\text{s}=\dfrac{2\text{sat}\left(K_\text{t}(t)(x(t)-l(t))\right)}{m_\text{s}}-\dfrac{\beta_\text{s}
(x_\text{s}(t))\dot{x}_\text{s}(t)}{m_\text{s}}-\dfrac{K_\text{s}x_\text{s}}{m_\text{s}},
\end{equation}
where $K_\text{s}$ is the spring's stiffness, $m_\text{s}$ the mass of the moving plate carrying the tether's pulley (we neglect the spring's mass) and $\beta_\text{s}(t)$ is the viscous friction coefficient of the mass-spring system. The latter is a function of $x_\text{s}$ to account for the end of travel of the spring in both directions:
\begin{equation}\label{E:spring_viscous}
\beta_\text{s}(t)=\left\{
\begin{array}{lcl}
\underline{\beta}_\text{s}(x_\text{s})&\text{if} &\text{ $\delta_\text{s}\leq x_\text{s}\leq\overline{x}_\text{s}-\delta_\text{s}$}\\
\gamma_\text{s}\underline{\beta}_\text{s}&\text{if} &\text{$0\leq x_\text{s}\leq\delta_\text{s}$ and $\dot{x}_\text{s}<0$}\\
 &&\text{or $x_\text{s}>\overline{x}_\text{s}-\delta$ and $\dot{x}_\text{s}>0$}
\end{array}
\right.,
\end{equation}
where $\overline{x}_\text{s}$ is the maximum spring travel, $\underline{\beta}_\text{s}$ is the friction coefficient when the spring is free to move (typically a very small value) and $\gamma_\text{s}\gg1$ is a coefficient accounting for the increase of friction when the spring is close to its travel limits, e.g. due to the presence of rubber bumpers. The travel limits are given by a small value $\delta_\text{s}>0$ when the spring is uncompressed, and $\overline{x}_\text{s}-\delta$ when it's fully compressed. Finally, we assume that the winch reels-out with the maximum available motor torque, $\overline{T}_\text{w}$. The equation of motion of the winch is:
\begin{equation}\label{E:winch_simple}
\ddot{\theta}_\text{w}=\dfrac{\overline{T}_\text{w}}{J_\text{w}}+\dfrac{R_\text{w}\,\text{sat}\left(K_\text{t}(t)(x(t)-l(t))\right)}
{J_\text{w}}-\dfrac{\beta_\text{w}\dot{\theta}_\text{w}(t)}{J_\text{w}},
\end{equation}
with $J_\text{w}$ being the winch moment of inertia and $\beta_\text{w}$ its rotational viscous friction coefficient. In order to simulate the model equations \eqref{E:tether_stiffness}-\eqref{E:winch_simple}, initial conditions for the states of the aircraft, spring and winch have to be set. In particular, we parameterize such initial condition as:
\begin{subequations}\label{E:simple_sim_init}
\begin{gather}
x(0)=x_0;\;\dot{x}(0)=\dot{x}_0\label{E:init_aircraft}\\
x_\text{s}(0)=0;\;\dot{x}_\text{s}(0)=0\label{E:init_spring}\\
\theta_\text{w}(0)=\dfrac{x_0}{R_\text{w}};\;\dot{\theta}_\text{w}(0)=\dfrac{\dot{x}(0)-\Delta\dot{x}}{R_\text{w}}\label{E:init_winch}\\
\end{gather}
\end{subequations}
where $x_0,\,\dot{x}_0$ are the initial position and speed of the aircraft, and $\Delta\dot{x}$ an initial difference of speed between the aircraft and the winch. The rationale behind \eqref{E:simple_sim_init} is the following: we assume that the aircraft is flying with a certain initial speed $\dot{x}_0$ and that the initial tether length is just enough to provide no tether force with the spring being at rest (compare equations \eqref{E:aircraft_simple}, \eqref{E:tether_length} and \eqref{E:init_spring}-\eqref{E:init_winch}), however with the winch speed lower than the aircraft's one by a quantity $\Delta\dot{x}$. When simulating this model, such a speed difference will induce a positive tether force which will slow down the aircraft and compress the spring, see eqs. \eqref{E:aircraft_simple} and \eqref{E:spring_simple}. At the same time, the winch will accelerate due to the applied torque and the tether pull, until at some time instant $t^*$ the tether length will be large enough to eventually bring the pulling force down to zero, this time with a winch speed equal or larger than the aircraft's. Then, we can evaluate whether the aircraft's speed in the time interval $[0,\;t^*]$ is always larger than the minimum cruise speed $\dot{\underline{x}}$, i.e. whether the aircraft is able to stay airborne. This test allows one to judge whether a given combination of spring's stiffness and maximum travel is suitable for the considered aircraft (mass, available thrust, drag coefficient, area), winch (inertia and available torque), and tether (stiffness) and for the selected initial aircraft speed $\dot{x}_0$ and speed difference $\Delta\dot{x}$. Such a simple simulation study can then be used to evaluate different design alternatives and to select one. Note that the described approach considers a worst-case scenario where the tether is perfectly aligned with the airspeed (hence with the drag force), while in reality the tether is pulling with an angle such that not all of its force adds to the drag.

\begin{figure}[!hbt]
 \begin{center}
  \includegraphics[width=\columnwidth]{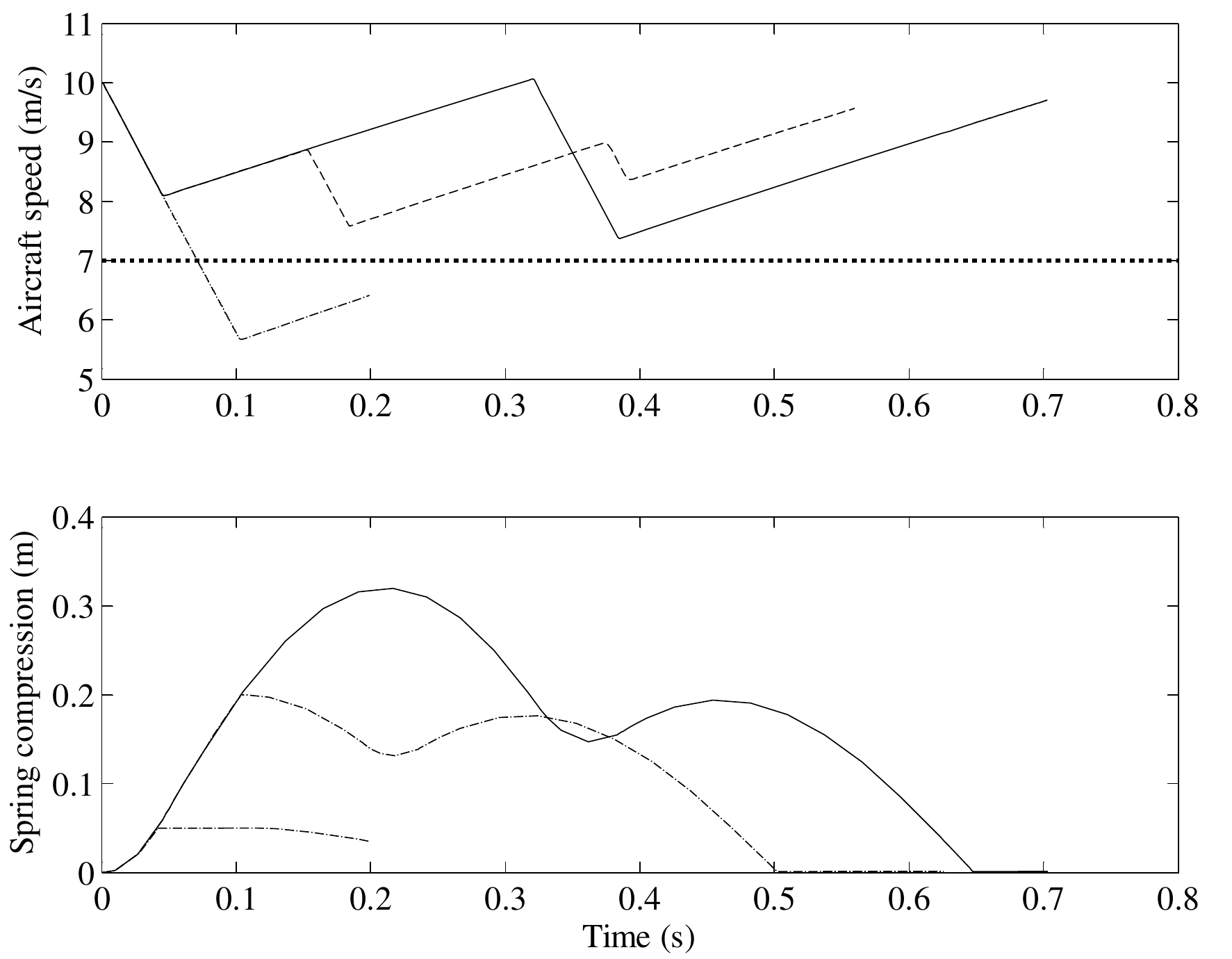}
  \caption{Simulation results obtained with the simple model of interaction between the winch, the mass-spring system, the tether and the aircraft. Comparison of aircraft speed (upper plot) and spring compression with $\overline{x}_\text{s}=0.35\,$m (solid line), 0.2$\,$m (dashed) and 0.05$\,$m (dash-dot). Dotted line: minimal allowed aircraft speed.}
  \label{F:spring_design_example}
 \end{center}
\end{figure}

An example of the mentioned design procedure pertaining to our small scale prototype is presented in Fig. \ref{F:spring_design_example}, where four different values of maximum spring compression (0.05m, 0.2m, and 0.35m) are compared, with all the other parameters reported in Table \ref{T:simple_sim_param}.

\begin{table}[!htb]
	\caption{Parameters of the simplified model used to design the mass-spring system}
	\label{T:simple_sim_param}
	\centering
	\begin{tabular}{|lrl|lrl|}\hline
\multicolumn{3}{|l|}{\textbf{Aircraft and ambient}}&\multicolumn{3}{l|}{\textbf{Mass-spring system}}\\\hline
$A$ & $0.3$ & m$^{2}$&$m_\text{s}$&$2$&kg\\\hline
$C_D$ & $0.05$ & -&$K_\text{s}$&$70$&N$\,$m\\\hline
$m$ & $1.2$ & kg&$\underline{\beta}_\text{s}$&$10^{-4}$&kg$\,$s$^{-1}$\\\hline
$\overline{T}$ & $10$ & N&$\gamma_\text{s}$&$10^6$&-\\\hline
$\dot{\underline{x}}$ & 7 & m$\,$s$^{-1}$&$\delta_\text{s}$&0.001&m\\\hline
$x_0$ & 20 & m&\multicolumn{3}{l|}{\textbf{Winch}}\\\hline
$\dot{x}_0$ & 10 & m$\,$s$^{-1}$&$R_\text{w}$&0.1&m\\\hline
$\rho$& $1.2$ & kg\,m$^{-3}$&$\overline{T}_\text{w}$&13&N$\,$m\\\hline
\multicolumn{3}{|l|}{\textbf{Tether}}&$J_\text{w}$&0.1&kg$\,$m$^2$\\\hline
$\overline{F}_\text{t}$&4500&N&$\beta_\text{w}$&0.01&kg$\,$m$^2\,$s$^{-1}$\\\hline
$\overline{\varepsilon}_\text{t}$&0.02&-&$\Delta\dot{x}$&4&m$\,$s$^{-1}$\\\hline
\end{tabular}
\end{table}

It can be noted that the spring with the smallest maximum travel does not provide enough buffer to keep the aircraft's speed above the minimum one. As the spring length increases, more than one cycle of compression and extension emerge and the minimum aircraft speed improves. Moreover, the time at which the minimum speed occurs is shifted farther in time as the available spring travel increases.

In our prototype, we used the described approach to design our spring system, whose maximum displacement is equal to 0.35$\,$m and all other parameters are equal to those shown in Table \ref{T:simple_sim_param}.

Finally, regarding the pulleys that redirect the tether from the winch to the aircraft, these have to be strong enough to withstand the involved forces, oppose little friction and be able to slightly adjust their orientation to cope with small misalignments and with the tether spooling. The pulleys' diameter shall be at least 20-30 times larger than the tether diameter to reduce wear. In our design we employed nylon stand-up pulleys used for sailing boats, see Fig. \ref{F:tensioning_system}, with a diameter of 0.045$\,$m.

\subsubsection{Power supply, motors and drives, measurement and control hardware and human-machine interface (HMI)}\label{SSS:drivetrain_design}
To maximize flexibility we designed a modular power supply, where the motor drives can be either connected to an external single-phase 220VAC source or to a 3kW inverter connected to 12V batteries with a total capacity of 260 Ah (see Fig. \ref{F:sys_concept}). We used the latter configuration in our outdoor tests. We sized the batteries to provide enough energy for at least one full day of testing, as well as large enough peak current. The latter can be estimated from the peak torque of the motors and their current/torque constant, considering the ratio between the voltage on the battery side and the one on the drive side.

Regarding the motors, since power production was not within the goals of our research we chose not to install a large machine for the winch. Rather we employed the same motor model, an ABB BSM90N-3150 permanent magnet servomotor, for both the slide and the winch. Each motor is equipped with incremental and Hall-effect encoders and it is connected to an ABB MicroFlex$^\circledR$ E150 servodrive. We chose the motor/drive combination according to quite standard design considerations as briefly outlined in the following. 
The motor rated torque is equal to $13\,$Nm, corresponding to a current of 9$\,$A, up to the rated speed of about 2000 rpm, which corresponds to about 21$\,$m/s of slide/tether speed, considering that the radius of both the slide drum and the winch is 0.1$\,$m. This value provides enough margin with respect to the take-off speed of the aircraft, approximately equal to 9 m/s. For short time intervals, the drives can provide twice the rated current to the motor (i.e. twice the torque): we employed this functionality for the slide motor in the very first acceleration at take-off. In particular, considering that the total mass of the slide drum, the slide and the aircraft is about 11.2$\,$kg and that the slide drum's radius is 0.1$\,$m, the peak torque of 26$\,$Nm can accelerate the aircraft from zero to take-off speed in about 0.38$\,$s using 1.7$\,$m of rails. This leaves enough room for the slide motor to break. The energy produced during the breaking of the slide and of the winch is dissipated in resistors connected to the drives, even though in principle one could  recover it. We programmed low-level speed and position control loops on the drives (see section \ref{S:control_design}) using an ABB basic-like language  called Mint$^\circledR$. We installed all the power supply components (batteries, inverter, a battery charger) and the drives in a metal box to protect them during road transportation.\\
As shown in Fig. \ref{F:sys_concept}, the drives send their position measurements to a real-time machine, which processes them together with the inputs received from the HMI and from the potentiometer measuring the spring's compression. The latter is a conductive plastic linear sensor (model LP-400F) manufactured by Midori Corp., with a maximum travel of 0.4$\,$m, which provides enough bandwidth and high resistance to outdoor environments (dust and water). As real-time machine we employed a Speedgoat$^\circledR$ performance machine in which we installed a National Instruments PCI-6221 data acquisition board. The real-time machine can be conveniently programmed in Matlab$^\circledR$/Simulink$^\circledR$ using rapid prototyping tools (Simulink Realtime$^\circledR$). Its main tasks are to compute the reference position and speed to be sent to the drives, and to log all the data related to the ground station.

Besides the development laptop used to modify the control programs running on the real-time machines and the drives, the HMI consists of a joystick to issue manually a position (resp. speed) reference to the slide (resp. winch) motor, and of switches used to enable/disable the drives and commence a take-off procedure.

\subsection{Aircraft}\label{SS:aircraft_design}
\subsubsection{Hardware and communication}\label{SSS:glider_hardware}
\begin{figure}[h]
	\centerline{
		\includegraphics[width=0.85\linewidth,clip]{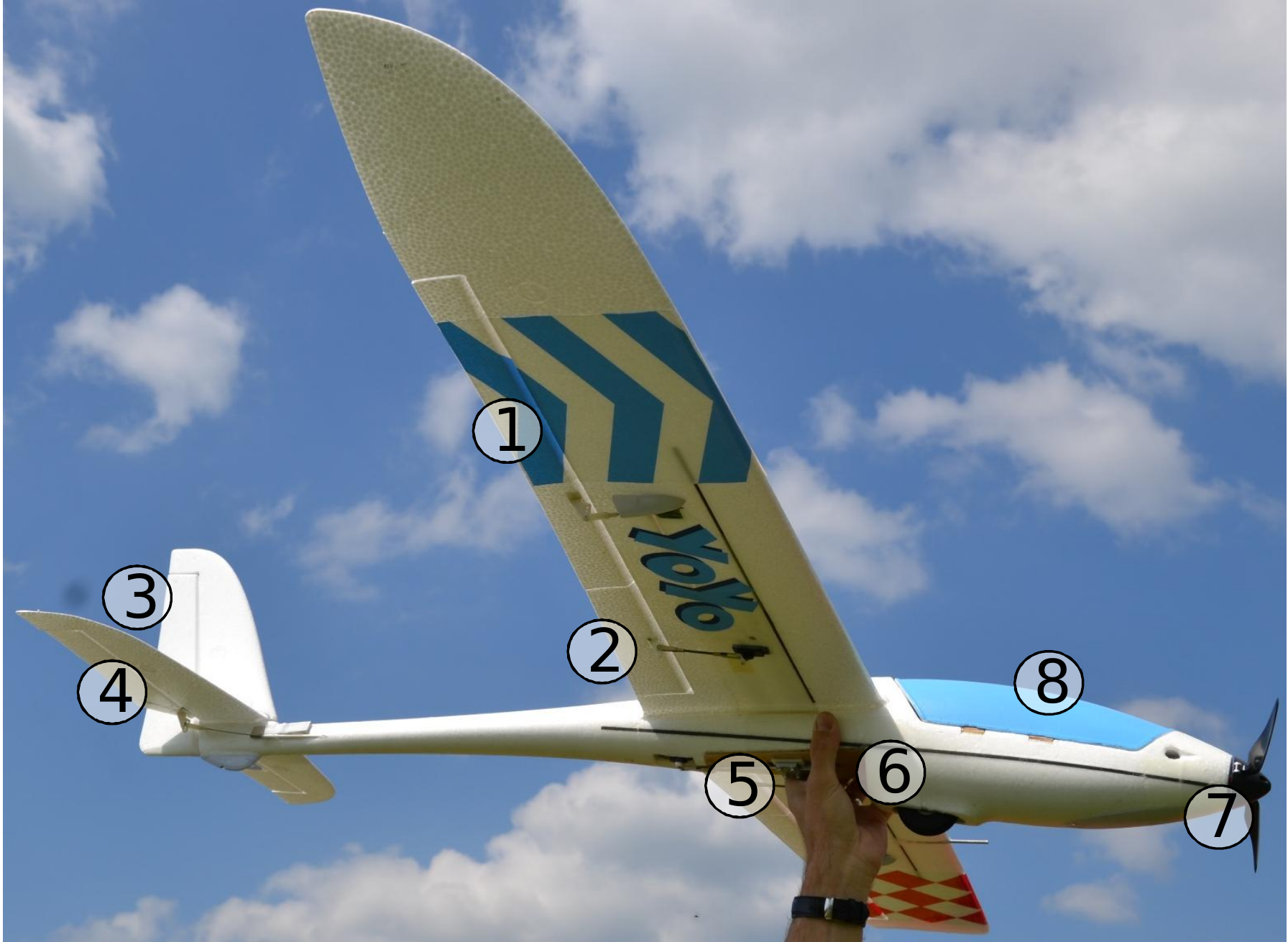}}
	\caption{Model glider used for the experiments. 1. ailerons; 2. flaps; 3.  rudder; 4. elevator; 5. release mechanism; 6. right-angle cut to engage with the slide; 7. propeller; 8. fuselage containing the measurement and control hardware (see Fig. \ref{F:Disassembled}).}\label{F:glider}
\end{figure}

We started from a commercially available glider that we modified for our needs, see Fig. \ref{F:glider}. 
%
In particular, we employed a model glider made out of styrofoam, whose advantages are resilience to impacts, low cost and the ease of structure modification to include additional hardware, while the main disadvantages are relatively large deformations of the structure during flight and high skin drag, which limit the flight speed. The chosen model glider also offers a large enough volume in its fuselage to host all the onboard components (sensors and control hardware). The geometrical characteristics of the glider are resumed in Table \ref{T:Geometry} and those of the propulsion system in Table \ref{T:pro}.

In order to connect the glider to the tether and be able to detach it during flight, we installed a release mechanism controlled by a servomotor and re-enforced the body with fiber glass and plywood. Such a release mechanism is  meant to be used to tow model gliders using a winch placed on ground, hence it is already designed to be mounted under the fuselage (like in our case) and not in the nose. Usually employed in competitions, these mechanisms can support high forces. We chose a model milled in aluminum and integrated it in the body close to the center of gravity.

To engage the glider on the slide, we cut the fuselage with a right angle, similarly to a hydroplane. In this manner, we obtain a horizontal surface, used to place the glider on the slide, and a vertical surface against which the slide pushes during the initial acceleration at take-off. In order to land the glider in a compact area, we added flaps.
\begin{table}[!htb]
	\caption{Main features of the employed model glider}
	\label{T:Geometry}
	\centering
	\begin{tabular}{|l|r|l|}\hline
		Wingspan & $1.68$& m\\\hline
		Wing surface area& $0.3174$&m$^2$\\\hline
		Aspect ratio& 8.89&-\\\hline
		Mean aerodynamic chord (MAC)& $0.194$&m\\\hline
		Mass & $1.2$&kg\\\hline
	\end{tabular}
\end{table}

For the communication between the glider and the pilot we used as radio link a 2.4GHz Futaba system composed of a T14SG remote control and an 6308SBT receiver. The remote control can transmit 14 channels and the receiver can control up to 12 servomotors using the S-bus protocol. This protocol is able to transmit all channels digitally through one cable instead of several dedicated pulse width modulation (PWM) lines. 

\begin{table}[h!]
	\caption{Main features of the glider propulsion system}
	\label{T:pro}
	\centering
	\begin{tabular}{|rrr|c|c|c|}\hline
		\multicolumn{3}{c|}{Brushless Motor} & \multirow{2}{*}{Driver} & \multirow{2}{*}{Propeller} & Lithium Polymer\\
		Model & Kv & A & & & battery\\
		\hline
		 D2836 & 1100& 20 & 30A& 9"$\times$6"& 14.4V, 1300mAH \\\hline
	\end{tabular}
\end{table}
Regarding the servomotors, we employed $9$g-class ones for all control surfaces (rudder, ailerons, elevator and flaps). In addition, the servomotor chosen to drive the release mechanism is a HS-82 MG with higher torque and reinforced gears. All servomotors have a travel of $\pm 90^\circ$ and are controlled in position by a PWM signal. The width of the pulse is generally between $1000\,\mu$s and $2000\,\mu$s and the frequency is 50$\,$Hz. The other features of the servos are resumed in Table \ref{T:Servo}.

\begin{table}[h!]
	\caption{Performance of the employed servomotors.}
	\label{T:Servo}
	\centering
	\begin{tabular}{|l|r|r|}\hline
		Servo Type & 9g & HS-82 MG \\
		\hline
		Weight ($10^{-3}\,$kg) & 9 & 19\\\hline
		Torque ($10^{-1}\,$Nm) & $15$ & $29$\\\hline
		Positioning time (s/60\degree) & $0.15$ & $0.12$ \\\hline
		Gears & Plastic & Metal\\\hline
	\end{tabular}
\end{table}

\subsubsection{Measurement and control}\label{SSS:glider_control}

\begin{figure}[h]
	\centerline{ \includegraphics[width=0.85\linewidth,clip]{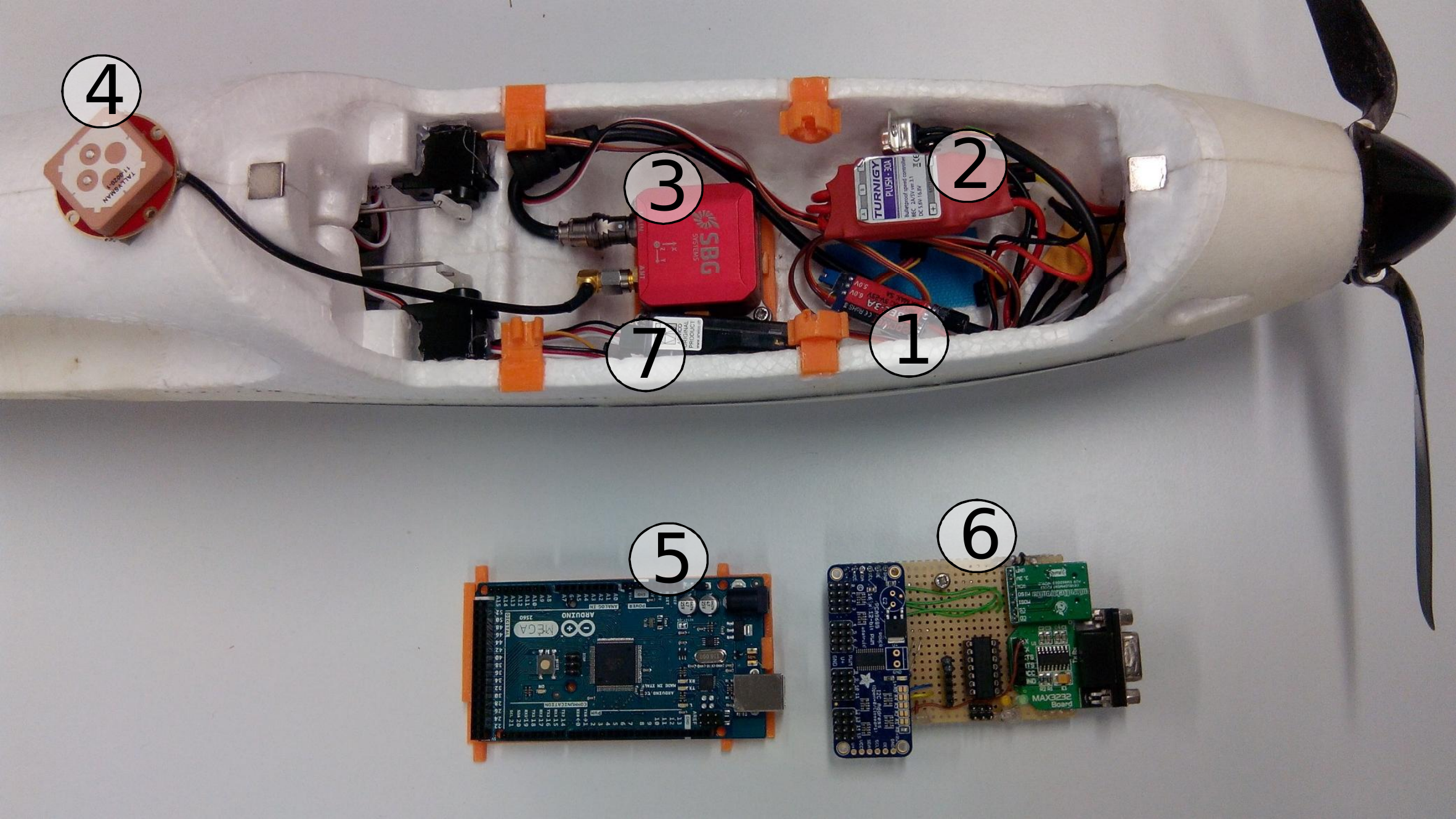}}
	\caption{Onboard measurement and control hardware for the glider. 1. DC-DC power converter; 2. motor driver; 3. inertial measurement unit; 4. GPS antenna; 5. Arduino$^\circledR$ Mega 2560 embedded platform; 6. extension board for RS-232 and S-bus interfaces; 7. RC receiver. The LiPo battery is not shown in the figure.}\label{F:Disassembled}
\end{figure}

\begin{figure}[!h!]
	\centerline{	 \includegraphics[width=0.85\linewidth,clip]{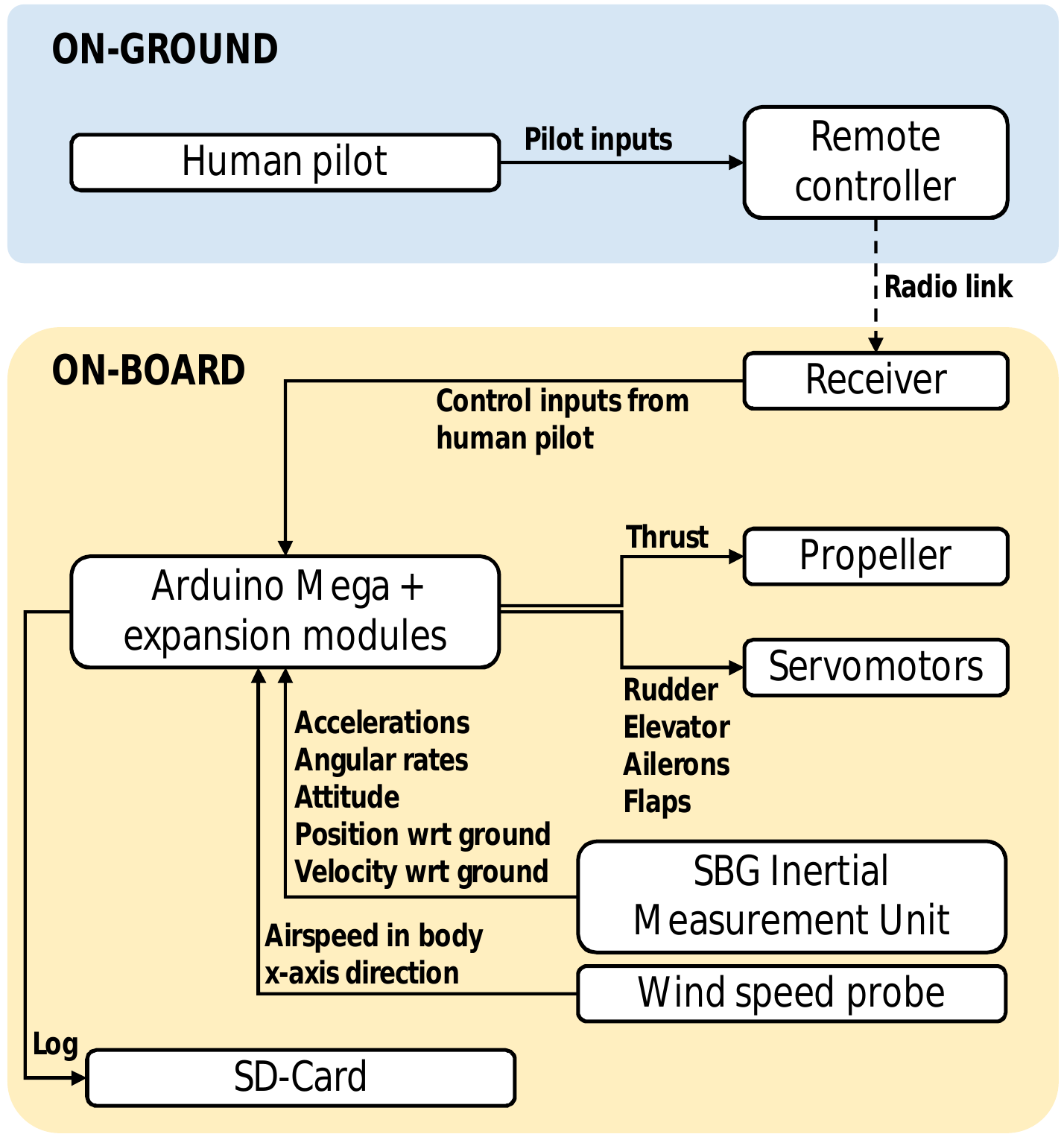}}
	\caption{Layout of the measurement and control system for the glider.}\label{F:glider_layout}
\end{figure}

We installed the measurement and control hardware inside the fuselage, except for the airspeed sensor which we attached to one of the wings. Fig. \ref{F:Disassembled} shows the main components. The conceptual layout of the glider's measurement and control system is depicted in Fig. \ref{F:glider_layout}. As main micro-controller, we employed an Arduino$^\circledR$ Mega 2560. The ease of programming, high number of input-output interfaces and the 5VDC operating voltage are the main reasons for this choice. The board acquires all signals from the sensors and pilot inputs, issues the control inputs to the actuators (control surfaces, release mechanism and motor) and logs data. We added an extension board (also shown in Fig. \ref{F:Disassembled}) to interface with the sensors and servomotors, supplying both power and control signals. The two boards together occupy a relatively large space and weigh about 0.2$\,$kg. Together with the sensors and the power sources, the overall weight of the measurement and control hardware is almost 0.3$\,$kg, which is a large share of the total weight (about 1.2$\,$kg) for such a small-scale, light glider. 

The attitude, angular velocity vector, accelerations, and inertial position and velocity vectors of the glider are measured by a commercial inertial measurement unit (IMU) by SBG Systems. The latter integrates inertial sensors (gyros and accelerometers), magnetometers, a barometer, and a GPS receiver. The IMU carries out sensor fusion with an extended kalman filter and provides a full state estimation at constant sampling frequency, 50$\,$Hz in our case. 
The unit weighs about 0.05$\,$kg plus about 0.02$\,$kg of the GPS antenna, which we installed on top of the fuselage, see Fig. \ref{F:Disassembled}. The antenna is a TW1421 model from Tallysman with a gain of 28dB.
%
For the airspeed sensor we employed a Pitot probe, installed on the wing far away from the motor, and a digital transducer (MS4525DO) that provides differential pressure measurements and temperature readings through an I2C protocol. The differential pressure has a range  $6894.76\,$Pa with an accuracy of $0.84$Pa and is acquired with a precision of 14$\,$bits. The temperature sensor has a range of -50$^\circ\,$C to 150$^\circ\,$C, with an accuracy of 1.5$^\circ\,$C and a precision of 11$\,$bits. The airspeed is then computed from the differential pressure and temperature readings, using the absolute pressure measurement given by the IMU.

Regarding the actuators, the servomotors are controlled by a PWM signal which we generate with the Adafruit Servo Driver Board based on a PCA9685 chip. The board communicates via I2C with Arduino and can command up to 16 servos. It has a 12$\,$bit precision which translates into a minimum step of $5.85\,\mu$s in the signal sent to the servo. Since most analog servos have a precision of $10\,\mu$s to $8\,\mu$s, one can obtain a very smooth control without jittering. During dynamic maneuvers, the servomotors can draw a significant amount of current. For this reason we connected the PWM board to two separated power supplies, one for the chip and the other for the servomotors: the chip is supplied by the main power source of the Arduino board, a DC-DC converter providing a regulated 5V, while the servos are supplied by the same DC-DC converter as the motor controller.

To log data, we used a micro-SD card slot from MikroElectronika interfacing via SPI with Arduino. We used an open-source library to rapidly write binary log on the SD-card. The logs can be decoded after each test session using e.g. Matlab. The logging function programmed on the Arduino board is synchronized with the IMU to log at 50Hz. 

When armed, the embedded controller always reads the incoming data from all the onboard sensors and the inputs from pilot through the receiver, it generates the control signals for the servos and the motor, and it logs data. A switch operated by the pilot changes the flight mode, from completely manual (i.e. the pilot inputs are fed directly to the actuators), to fly-by-wire (with low-level controllers that stabilize the glider pitch and roll, and the pilot providing references for these angles), to fully autonomous. The modeling and automatic control of the glider is described in detail in a separate contribution \cite{Fagiano2016submitteda}.
%

\section{Ground station control design}\label{S:control_design}
The control structure for the ground station is hierarchical: inner feedback control loops programmed on the drives track references of position (for the slide) and speed (for the winch), which are computed by an outer control loop programmed on the real-time machine. In the following we describe in detail the control algorithms, while we provide the numerical values of the involved parameters in section \ref{S:results}. All control loops are implemented in discrete time with sampling period $T_s$; we denote with $k\in\mathbb{Z}$ the discrete time instants.

For the inner loops we employ static state-feedback controllers designed via pole-placement, using a standard linear model of the motors which takes into account inertia and viscous friction. Denoting with the subscript ``$_\text{s}$'' and ``$_\text{w}$'' quantities related to the slide and the winch motors, respectively, the corresponding control laws are:
\begin{subequations}\label{E:inner_loops}
\begin{gather}
u_{\text{s}}(k)=K_{\theta,\text{s}}(\theta_\text{ref,s}(k)-\theta_{\text{s}}(k))-K_{\dot{\theta},\text{s}}\,\dot{\theta}_{\text{s}}(k)\label{E:pos_control_slide}\\
u_{\text{w}}(k)=K_{\dot{\theta},\text{w}}(\dot{\theta}_\text{ref,w}(k)-\dot{\theta}_{\text{w}}(k))\label{E:spd_control_winch}\\
\end{gather}
\end{subequations}
where $u_{\text{s}},\,u_{\text{w}}$ are the commanded torques, $\theta_{\text{s}},\,\theta_{\text{w}}$ the angular positions of the motors, $\theta_\text{ref,$\text{s}$}$ the reference position for the slide motor, $\dot{\theta}_\text{ref,\text{w}}$ the reference speed for the winch motor, and $K_{\theta,\text{s}},\,K_{\dot{\theta},\text{s}},\,K_{\dot{\theta},\text{w}}$ the feedback gains. The state variables $\theta_{\text{s}},\,\dot{\theta}_{\text{s}},\,\theta_{\text{w}},\,\dot{\theta}_{\text{w}}$ are measured accurately with the encoders (incremental and Hall effect) mounted on the motors and connected to the drives. The torques are saturated to their maximum limits $\pm\overline{T}_{\text{s}},\,\pm\overline{T}_{\text{w}}$.

\begin{figure}[!h]
	\centerline{ \includegraphics[width=\linewidth,clip]{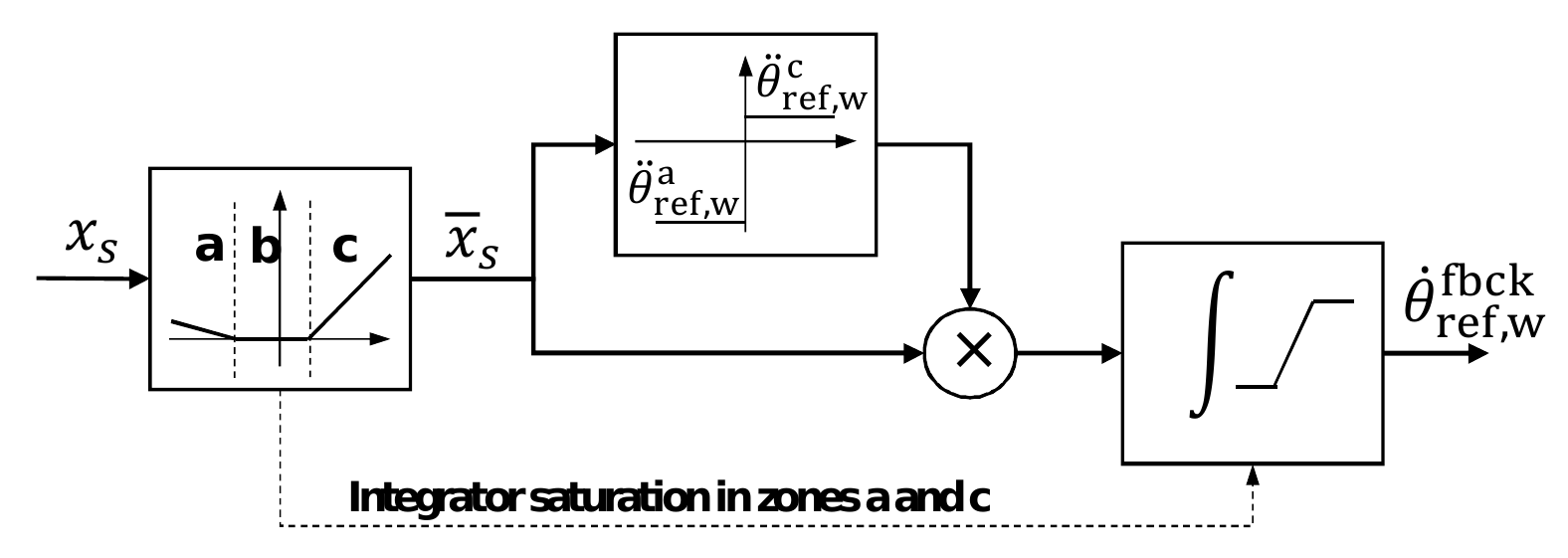}}
	\caption{Block-diagram of the feedback contribution to the reference winch speed.}\label{F:winch_control_sketch}
\end{figure}

For the outer loops, we implemented two modes: manual operation and automatic one. In the first mode, the references $\theta_\text{ref,s},\,\dot{\theta}_\text{ref,w}$ are issued by a human operator using the joystick in the HMI. This operation mode is used for debugging and to collect data for parameter identification (e.g. to estimate the inertia of the slide motion system and of the winch).\\
To carry-out a take-off maneuver, the second operating mode is used instead. The starting conditions are with the glider installed on the slide, and the latter positioned at one end of the rails. Without loss of generality let us assume that $k=0$ when the take-off maneuver starts. Then, a step of position reference $\theta_\text{ref,\text{s}}(k)=L/R_{\text{s}},\,\forall k>0$ is issued, where $L$ is the desired slide travel used for take-off and $R_{\text{s}}$ the radius of the slide drum. As a consequence, the inner controller for the slide will move it as fast as possible to the new reference position and during this motion the take-off speed of the aircraft is reached. The winch has to coordinate its motion with the slide and always achieve a good tradeoff between limiting the pulling force exerted by the tether, and avoiding tether entanglement, which quickly occurs when reeling-out without any load. In order to achieve this goal throughout the take-off and during flight, we employ a combined feedforward/feedback approach to compute the reference winch speed. The feedforward contribution latches the winch speed to the slide speed:
\begin{equation}\label{E:ffwd_winch}
\dot{\theta}_\text{ref,w}^\text{ffwd}(k)=\gamma\dot{\theta}_\text{s}(k)
\end{equation}
where $\gamma>0$ is a scalar that can be tuned to achieve a good coordination between the two drums. In our prototype, since the presence of the glider does not affect much the inertia of the slide and of the winch (due to their relatively much larger mass), the tuning of $\gamma$ could be carried out in preliminary laboratory tests without the aircraft. The feedback contribution exploits the measure of the spring travel $x_\text{s}(k)$. In particular, we set two threshold values, $x_\text{s}^I,\,x_\text{s}^{II}$, which divide the available spring travel in three zones:
\begin{itemize}
  \item \textbf{Zone a} ($0\leq x_\text{s}(k)\leq x_\text{s}^I$): the spring is practically uncompressed, the winch shall decrease speed and eventually reel-in;
  \item \textbf{Zone b} ($x_\text{s}^I < x_\text{s}(k)<x_\text{s}^{II}$): the spring is subject to low force, the winch shall be held in place (constant tether length);
  \item \textbf{Zone c}: ($x_\text{s}^{II}\leq x_\text{s}(k) \leq \overline{x}_\text{s} $): the spring is subject to relatively large force, the winch shall increase its speed and reel-out to release the tether.
\end{itemize}
Then, the feedback contribution to the reference winch speed is computed according to the following strategy (see Fig. \ref{F:winch_control_sketch}): \small
\begin{equation}\label{E:fbck_winch}
\begin{array}{l}
\verb"If " 0\leq x_\text{s}(k)<x_\text{s}^I \text{ (\textbf{Zone a})}\\
\\
\;\;\bar{x}_\text{s}(k)=\dfrac{x_\text{s}(k)-x_\text{s}^I}{x_\text{s}^{I,\text{a}}-x_\text{s}^I}\\
\;\;\dot{\theta}_\text{ref,w}^\text{fbck}(k)=\min\left(0,\max\left(\underline{\dot{\theta}}_\text{ref,w}^\text{fbck},\,
\left(\dot{\theta}_\text{ref,w}^\text{fbck}(k-1)+T_s\ddot{\theta}_\text{ref,w}^\text{a}\bar{x}_\text{s}(k)\right)\right)\right)\\
\\
\verb"Else if " x_\text{s}^I \leq x_\text{s}(k)<x_\text{s}^{II} \text{ (\textbf{Zone b})}\\
\\
\;\;\dot{\theta}_\text{ref,w}^\text{fbck}(k)=\dot{\theta}_\text{ref,w}^\text{fbck}(k-1)\\
\\
\verb"Else if " x_\text{s}^{II}\leq x_\text{s}(k) \leq \overline{x}_\text{s} \text{ (\textbf{Zone c})}\\
\\
\;\;\bar{x}_\text{s}(k)=\dfrac{x_\text{s}(k)-x_\text{s}^{II}}{x_\text{s}^{II,\text{c}}-x_\text{s}^{II}}\\
\;\;\dot{\theta}_\text{ref,w}^\text{fbck}(k)=\max\left(0,\min\left(\overline{\dot{\theta}}_\text{ref,w}^\text{fbck},\,
\left(\dot{\theta}_\text{ref,w}^\text{fbck}(k-1)+T_s\ddot{\theta}_\text{ref,w}^\text{c}\bar{x}_\text{s}(k)\right)\right)\right)\\
\end{array}
\end{equation}\normalsize
where $\underline{\dot{\theta}}_\text{ref,w}^\text{fbck},\,\overline{\dot{\theta}}_\text{ref,w}^\text{fbck}$ are the desired minimum and maximum reference speed values that can be issued, and $\ddot{\theta}_\text{ref,w}^\text{a},\,\ddot{\theta}_\text{ref,w}^\text{c}$ are the desired angular accelerations for the reference speed. Such values are scaled according to the position of the potentiometer relative to the values $x_\text{s}^{I,\text{a}}<x_\text{s}^I$ and $x_\text{s}^{II,\text{c}}>x_\text{s}^{II}$, respectively for zones \textbf{a} and \textbf{c}, which are design parameters as well. Equation \eqref{E:fbck_winch} represents in practice an integral controller, where the integrated quantity is the distance of the spring position $x_\text{s}(k)$ from the interval $(x_\text{s}^I,\,x_\text{s}^{II})$ (i.e. zone \textbf{b}) and the gain is piecewise constant, since it is different in zones \textbf{a} and \textbf{c}. Moreover, a saturation of the integrated variable to negative (resp. positive) values is operated whenever the spring enters zone \textbf{a} (resp. \textbf{c}), in order to quickly start to reel-in (resp. reel-out) when the tether is released (resp. pulled).  As shown in the next section, a sensible choice for the involved design parameters is $\underline{\dot{\theta}}_\text{ref,w}^\text{fbck},\,\ddot{\theta}_\text{ref,w}^\text{a}<0$, $\overline{\dot{\theta}}_\text{ref,w}^\text{fbck},\,\ddot{\theta}_\text{ref,w}^\text{c}>0$, $x_\text{s}^{I,c}\approx x_\text{s}^I/2$ and $x_\text{s}^{I,c}\approx (\overline{x}_\text{s}+x_\text{s}^{II})/2$.

Finally, the reference speed actually issued to the low-level controller is computed as:
\begin{equation}\label{E:ref_spd_winch}
\begin{array}{l}
\verb"If " \dot{\theta}_\text{s}(k)>0\\
\\
\;\;\dot{\theta}_\text{ref,w}(k)=\max\left(\dot{\theta}_\text{ref,w}^\text{ffwd}(k),\dot{\theta}_\text{ref,w}^\text{fbck}(k)\right)\\
\\
\verb"Else "\\
\\
\;\;\dot{\theta}_\text{ref,w}(k)=\dot{\theta}_\text{ref,w}^\text{fbck}(k)
\end{array}
\end{equation}
According to \eqref{E:ref_spd_winch}, the feedforward contribution is used only if larger than the feedback one, and only if the speed of the slide is positive, i.e. during take-off.

\section{Experimental results}\label{S:results}
Table \ref{T:control_param} shows the values of the controller parameters that we used in our experimental tests. The system parameters (ground station and glider) are the ones reported in Tables \ref{T:simple_sim_param}-\ref{T:Servo} and introduced throughout the paper. The chosen spring has a maximum travel $\overline{x}_\text{s}=0.35\,$m.
\begin{table}[!htb]
	\caption{Control parameters used for the experiments}
	\label{T:control_param}
	\centering
	\begin{tabular}{|lrl|lrl|}\hline
$T_s$ & $0.001$ & s&$x_\text{s}^{II}$& $0.1$ & m\\\hline
$K_{\theta,\text{s}}$ & $14$ & Nm$\,$rad$^{-1}$&$x_\text{s}^{I,c}$& $0.025$ & m\\\hline
$K_{\dot{\theta},\text{s}}$ & $2.5$ & Nm$\,$s$\,$rad$^{-1}$ &$x_\text{s}^{II,c}$& $0.2$ & m\\\hline
$K_{\dot{\theta},\text{w}}$ & $1$ & Nm$\,$s$\,$rad$^{-1}$ &$\underline{\dot{\theta}}_\text{ref,w}^\text{fbck}$& $-10$ & rad$\,$s$^{-1}$\\\hline
$\overline{T}_{\text{s}}$ & $26$ &Nm&$\overline{\dot{\theta}}_\text{ref,w}^\text{fbck}$ & $120$ & rad$\,$s$^{-1}$\\\hline
$\overline{T}_{\text{w}}$ & $13$ & Nm&$\ddot{\theta}_\text{ref,w}^\text{a}$& $-100$ & rad$\,$s$^{-2}$\\\hline
$\gamma$ & $1.2$ & -& $\ddot{\theta}_\text{ref,w}^\text{c}$& $30$ & rad$\,$s$^{-2}$\\\hline
$x_\text{s}^I$& $0.05$ & m& L& 3.7 &m\\\hline
	\end{tabular}
\end{table}

\begin{figure}[!h]
	\centerline{ \includegraphics[width=0.9\linewidth,clip]{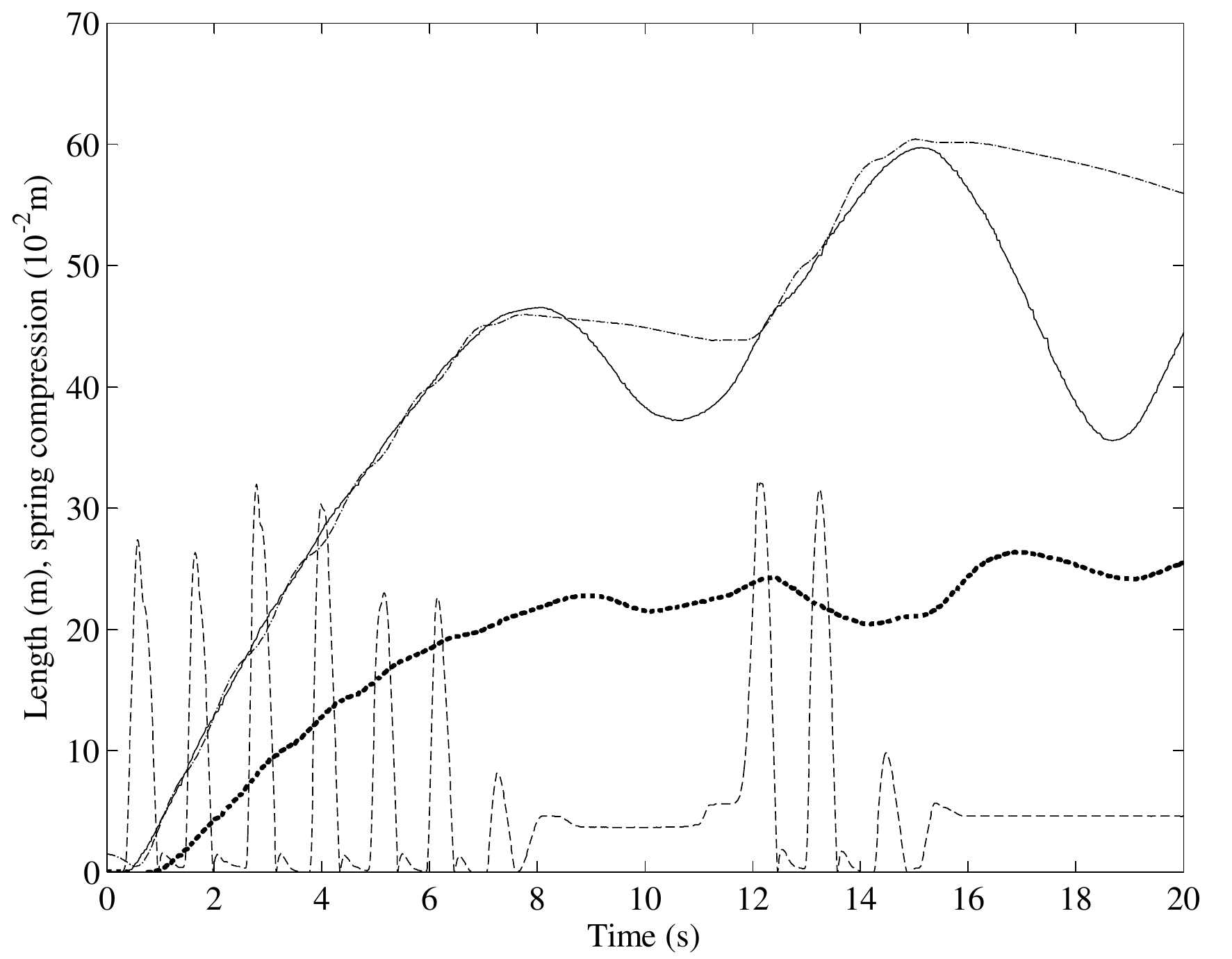}}
	\caption{Experimental results. Courses of the glider elevation (thick dotted line), its distance from the ground unit (solid line), the tether length between the glider and the slide (dash-dot), and the spring compression (dashed). The latter is expressed in $10^{-2}\,$m for the sake of readability, all other quantities are in m.}\label{F:glider_elev}
\end{figure}

We present here the typical results obtained during a take-off test with manually piloted glider and autonomous ground station. Fig. \ref{F:glider_elev} shows the elevation of the glider above ground during the take-off, together with its distance from the ground station, the tether length measured between the slide and the glider (i.e. taking into account also the slide motion), and the spring compression.
\begin{figure}[!h]
	\centerline{ \includegraphics[width=0.9\linewidth,clip]{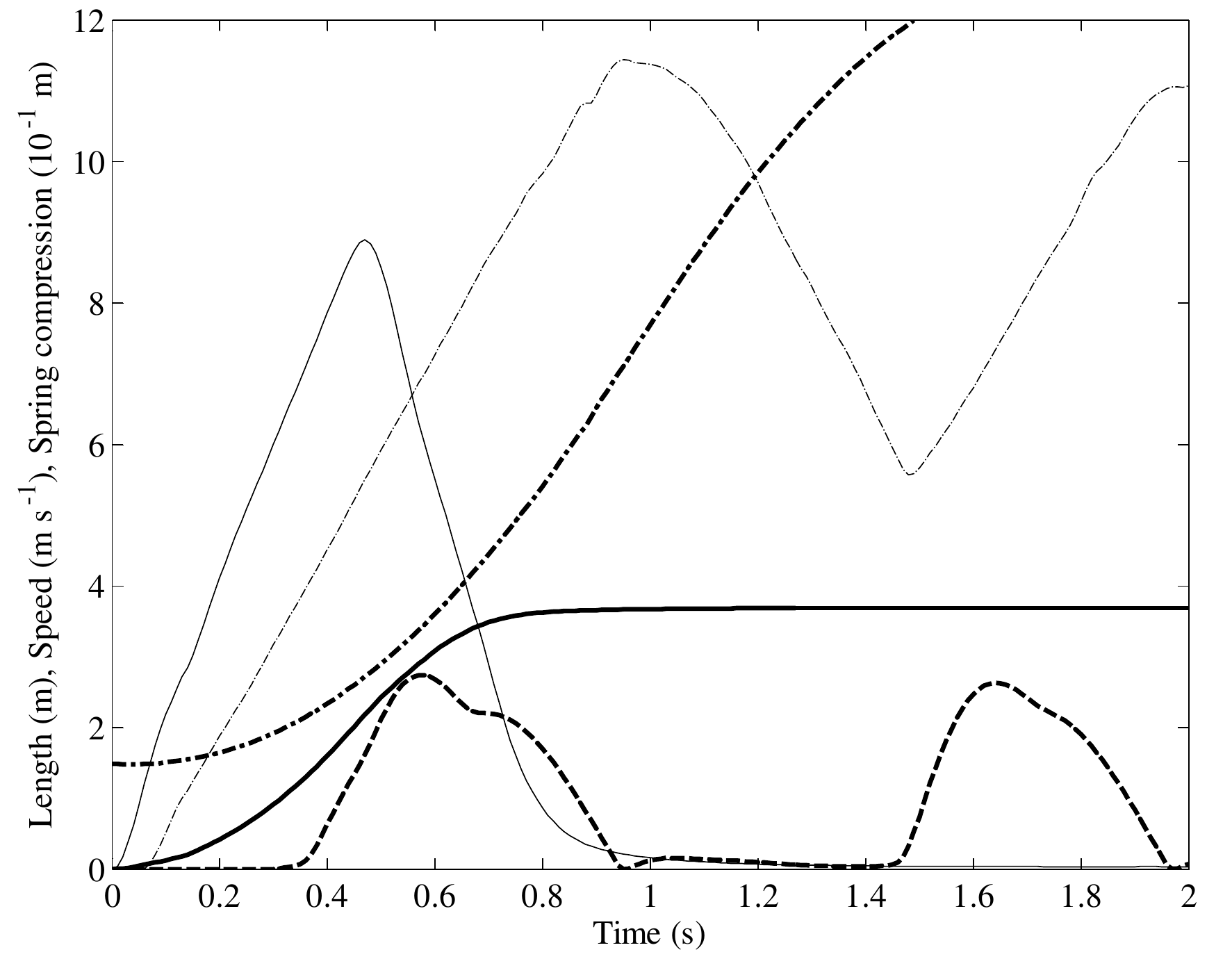}}
	\caption{Experimental results. Courses of the slide position (thick solid line), winch position (thick dash-dotted line) in m,  spring compression (thick dashed line) in $10^{-1}\,$m, and of the slide speed (solid line) and winch speed (dash-dotted line) in m$\,$s$^{-1}$.}\label{F:ground_station}
\end{figure}
\begin{figure}[!h]
	\centerline{ \includegraphics[width=0.9\linewidth,clip]{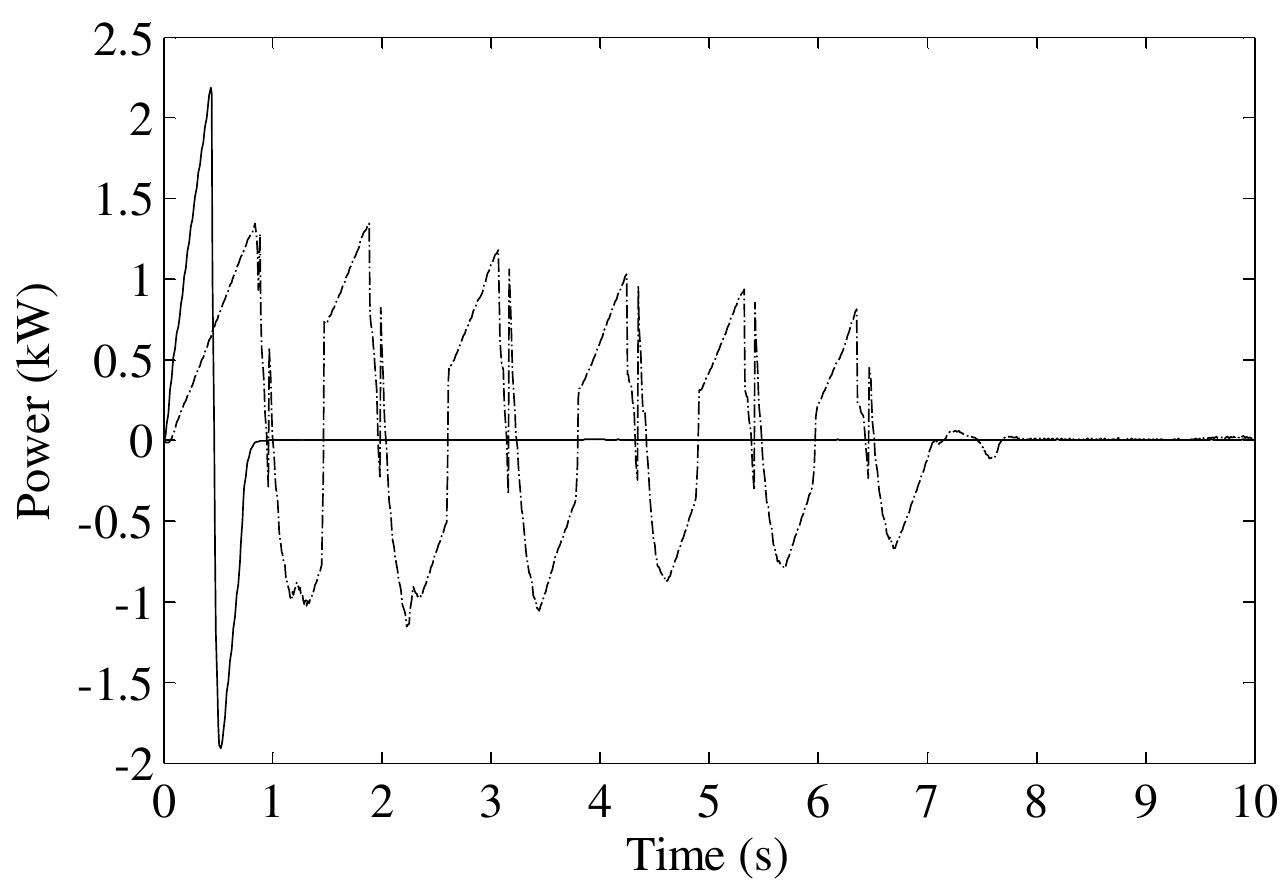}}
	\caption{Experimental results. Courses of the power consumed by the slide motor (solid line) and by the winch motor (dash-dotted line).}\label{F:power}
\end{figure}
The corresponding average climb angle up to 20$\,$m of elevation is about 30$^\circ$. It can be noted that a very good agreement exist between the spring movement and the difference between the glider distance from the ground station, measured by the onboard IMU, and the tether length, measured with the winch motor encoder. Indeed, these two measurements allow one to obtain a quite accurate estimate of the length of slack tether. The behavior of the slide and of the winch during the first few seconds of the same test is presented in Fig. \ref{F:ground_station}, where all quantities are expressed in m and m/s for the sake of comparison. The lag between the speed of the winch and that of the slide is due to the higher inertia of the former (and lower applied maximum torque, see Table \ref{T:control_param}); we compensated the resulting position difference by leaving a small initial slack line length, visible in Fig. \ref{F:ground_station} in terms of non-zero winch position at time $t=0$. Fig. \ref{F:ground_station} also illustrates the typical behavior of the control strategy given by \eqref{E:ffwd_winch}-\eqref{E:ref_spd_winch}: during the very first time instants the winch speed is increasing even if the spring is uncompressed, since the feedforward contribution is larger than the feedback one, while after about 0.5$\,$s the feedback strategy based on the spring compression is used. Overall, the described hardware solution and the corresponding control approach achieve the desired, conflicting goals of limiting the tether pull while avoiding an excessive reel-out of the line, which would lead to entanglement very quickly. A movie showing the winch behavior (in combination with the glider autopilot described in detail in \cite{Fagiano2016submitteda}) is available on-line \cite{Fagiano2016a}.

Finally, Fig. \ref{F:power} shows the courses of the power consumed by the winch and the slide during take-off. The negative power (i.e. power required to brake the motors) is dissipated in the braking resistors connected to the drives in our prototype, while in a final solution they could be recovered with storage or fed back to the grid. The peak power values are 2.18$\,$kW and 1.34$\,$kW respectively for the slide and winch motor. The theoretical analysis of \cite{Fagianoa} predicts, for the same glider and ground station features, values of 2.11 and 1.26 kW, very close to the ones obtained in our experiments, hence confirming the validity of the results in \cite{Fagianoa} pertaining to this take-off approach.

\section{Conclusions}\label{S:conclusions}
The paper presented in detail the design of a small-scale prototype to study the take-off of tethered aircrafts. Such a design can be easily replicated and improved by researchers and developers working on airborne wind energy systems with ground-based generation and rigid wings. The reported experimental results show that a quite effective take-off maneuver can be achieved in compact space, with power requirements in line with previous theoretical findings. The sensible next steps along this line of research are a similar study for other types of take-off strategies, e.g. using vertical-axis propellers, in order to further validate the existing theoretical analyses \cite{Fagianoa}, the development of landing strategies (including their experimental validation) with similarly effective performance in terms of low cost and compactness, finally the study of full operational cycles of take-off, power generation, and landing.


%

%

\section*{Acknowledgment}
The authors would like to thank Alessandro Lauriola and Stefan Schmidt for their helpful contributions during the project.

\bibliographystyle{plain}

\end{document}